\documentclass[nofootinbib,apsspec,showpacs,twocolumn,groupedaddress,lengthcheck,preprintnumbers]{revtex4}
\usepackage{amsmath}
\usepackage{amssymb}
\usepackage{graphics}
\usepackage[dvips]{graphicx}
\usepackage{graphicx}%
\usepackage[usenames]{color}

\def\simge{\mathrel{%
       \rlap{\raise 0.511ex \hbox{$>$}}{\lower 0.511ex \hbox{$\sim$}}}}
\def\simle{\mathrel{
       \rlap{\raise 0.511ex \hbox{$<$}}{\lower 0.511ex \hbox{$\sim$}}}}

\newcommand{\angstrom}{\mbox{\normalfont\AA}}
\newcommand \beq{\begin{eqnarray}}
\newcommand \eeq{\end{eqnarray}}

\newcommand{\redflag}[1]{{\color{red} #1}}

\usepackage[colorlinks=true,linktocpage=true,linkcolor=blue,citecolor=blue]{hyperref}
\usepackage{hyperref}

\begin{document}
\title{Searching for low mass dark matter via phonon creation in superfluid $^4$He}
\author{Gordon Baym,$^{a,b}$ D.\ H.\ Beck,$^a$ Jeffrey P. Filippini,$^a$ C.\ J.\  Pethick,$^{a,b,c}$ and Jessie Shelton$^a$}
\affiliation{\mbox{$^a$Illinois Center for Advanced Studies of the Universe \& Department of Physics,}\\
\mbox{University of Illinois at Urbana-Champaign, Urbana, IL 61801} \\
\mbox{$^b$The Niels Bohr International Academy, The Niels Bohr Institute, University of Copenhagen,}\\
\mbox{Blegdamsvej 17, DK-2100 Copenhagen \O, Denmark}\\	
\mbox{$^c$NORDITA, KTH Royal Institute of Technology and Stockholm University,}\\
\mbox {Roslagstullsbacken 23, SE-10691 Stockholm, Sweden} 
}

\date{\today}
\begin{abstract}
 
 We consider the scattering of dark matter particles from superfluid liquid $^4$He, which has been proposed as a target for their direct detection.  
Focusing on dark matter masses below $\sim$ 1 MeV, we demonstrate from sum-rule arguments the importance of the production of single phonons with energies $\omega \lesssim 1$ meV.
 We show further that the anomalous dispersion of phonons in liquid $^4$He at low pressures [i.e., $d^2\omega(q)/dq^2>0$, where $q$ and $\omega(q)$ are the phonon momentum and energy] has the important consequence that a single phonon will decay over a relatively short distance into a shower of lower energy phonons centered on the direction of the original phonon.   Thus the experimental challenge in this regime is to detect a shower of low energy phonons, not just a single phonon.   Additional information from the distribution of phonons in such a shower could enhance the determination of the dark matter mass.
\end{abstract}

\maketitle

\section{Introduction}

  Although the existence of dark matter has been conclusively established by multiple independent lines of gravitational evidence~\cite{Bertone:2016nfn}, its nature remains one of the outstanding mysteries in physics. Weakly Interacting Massive Particle (WIMP) models of dark matter, which rely on Standard Model interactions to produce the dark matter relic abundance, have thus been an important experimental target for decades~\cite{Goodman:1984dc,Jungman:1995df}.   A broad range of experiments has been deployed to directly detect the elastic scattering of 
  $\sim$ 10 GeV-mass WIMPs from heavy nuclei (see, e.g., \cite{Schumann:2019eaa} and references therein).
  
  With WIMP dark matter now under increasing observational strain, broadening the scope of terrestrial searches for dark matter has become ever more compelling \cite{Alexander:2016aln,Battaglieri:2017aum}.  
If dark matter interacts with the matter of the Standard Model via a new, dark force, the mass range of interest for direct detection experiments becomes much broader, and in particular extends down to the observational warm dark matter limit of order a few keV~\cite{Viel:2013apy}.  The direct detection of sub-MeV dark matter poses substantial challenges, due to the poor kinematic match with atomic nuclei and the very low available kinetic energy, $<1$~eV for sub-MeV dark matter moving at typical galactic velocities ($v/c\sim 10^{-3}$). Many interesting new or proposed experiments aim at dark matter masses $m_\chi$ in the MeV-to-a-few-GeV range, using either electronic scattering \cite{Essig:2011nj} in a variety of systems such as  semiconductors \cite{Graham:2012su, Essig:2015cda, Agnese:2018col, Abramoff:2019dfb, Aguilar-Arevalo:2019wdi}, liquid noble gases \cite{Essig:2017kqs, Agnes:2018oej,Aprile:2019xxb}, and other materials \cite{Derenzo:2016fse, Blanco:2019lrf}, or new channels to observe nuclear scattering \cite{Budnik:2017sbu,Essig:2019kfe,Baxter:2019pnz,Essig:2019xkx}. Far fewer experiments have been proposed to detect dark matter in the challenging sub-MeV regime.  Such schemes generally involve systems with very low energy gapped excitations, e.g., quasiparticles in superconductors~\cite{Hochberg:2015pha}, electrons in Dirac materials~\cite{Hochberg:2017wce},  and optical phonons in polar crystals \cite{Knapen:2017ekk}.  More recently, Ref.~\cite{Griffin} has called attention to the advantages of using materials with high sound speeds.

   Superfluid $^4$He is a particularly promising target for the detection of light dark matter particles. Atomic helium recoils from GeV-mass particles can be detected via the resulting electronic excitations, visible as scintillation and ionization~\cite{guo}.  Lighter particles can excite phonons and rotons, the meV-scale collective excitations of the superfluid, as discussed in detail by Schutz and Zurek~\cite{zurek} and by Knapen, Lin, and Zurek~\cite{zurek1}.   Such excitations may further evaporate individual $^4$He atoms from the superfluid surface, forming the basis of a detection scheme proposed by Maris et al. \cite{maris-seidel,ito-seidel} and recently explored by Hertel et al. \cite{hertel}.
  
   In this paper we explore in detail the physics of excitations produced by the scattering of dark matter particles in superfluid helium, focusing on the challenging mass range $m_{\chi}\lesssim 1$~MeV.   Such dark matter particles are not energetic enough to excite helium atoms electronically. 
    We formulate the interaction between the dark matter particle $\chi$ and $^4$He atoms in terms of a low energy s-wave pseudopotential -- essentially a contact interaction.    Predicting the interaction rate then reduces to understanding the density fluctuations in the helium, which at low energy are single phonons and rotons, as well as multiple phonon and roton excitations. Through the use of sum rules and explicit calculations, we constrain the production and damping of these excitations across a wide kinematic range.   Our focus throughout is on deriving the response of the superfluid, rather than on proposing a detector design or dark matter model. 
        
     The present analysis extends that of Zurek et al. \cite{zurek,zurek1} in two significant ways.  First, these authors limited themselves to processes that could generate phonons with energies above 1~meV, an assumed detection limit.    Since the maximum phonon-roton energy (the {\em maxon}) is $\sim$1.1~meV, this cut effectively excludes single phonon processes, and requires 
multiphonon excitations.  They draw upon theoretical calculations of high frequency density fluctuations \cite{Campbell} in estimating detection rates.
Here we consider the generation and propagation of excitations over a broader range of excitation energies, leaving aside for the moment issues of detectability.
   As we discuss more fully below, the f-sum rule for phonon fluctuations implies that as the phonon momentum decreases, single phonon processes become more and more dominant, exhausting some 90\% of the allowed weight even at the highest momentum transfer.

 A second feature we take into account here is the important role played by anomalous dispersion: the slight deviation from linearity of the low-energy phonon spectrum.  
Anomalous dispersion allows single low momentum phonons to decay into two (or more) phonons nearly collinear with the initial phonon -- the Beliaev process \cite{spartak,maris}.   As we show in detail, this process leads to rapid formation of phonon cones, analogues of cosmic ray air showers in the atmosphere, after the creation of a single phonon.\footnote{Acanfora et al.~present an effective field theory approach to the problem of detecting sub-GeV dark matter in superfluid $^4$He \cite{Acanfora}, in which the phonon dispersion relation is purely linear.  While parts of that discussion parallel the treatment here, their approach does not account for the physics of anomalous dispersion. See also Ref.~\cite{Caputo,Caputo2}.}   Although the detection of such soft phonon cascades is extremely challenging, their shape and extent at the helium surface encodes information on the location of the initial interaction and momentum direction beyond that available from a calorimetric measurement of the initial phonon.

This paper is organized as follows.
 In the next section we introduce the cast of characters:  the dark matter halo in the neighborhood of the Earth, and the excitations of superfluid $^4$He.  We then, in Sec.~III, review the kinematics of the interaction between dark matter particles and the helium, and model their interaction  in terms of a low energy pseudopotential.   In the following section, IV, we show that the f-sum rule bounds the rate of multi-excitation emission compared with the single phonon rate to at most 10\% at $q\lesssim$ 0.35 \AA$^{-1}$, determine the rate of single phonon emission, and describe phonon splitting and damping in the anomalous dispersion regime.  We discuss phonon damping in Sec.~V, two phonon production in Sec.~VI, and turn in the following section VII to describing the phonon cascade produced by an initial single phonon of low $q$, and outline how detection of such a cascade would proceed.   Appendix A is devoted to a technical discussion of the relation of the helium structure function, $S(q,\omega)$, and the helium density-density correlation function, Appendix B discusses $S(q,\omega)$ at non-zero temperature, and Appendix C discusses the $q$ dependence of the rate of direct production of a pair of phonons.

\section{Physical setting}

  The problem of dark matter scattering in superfluid helium lies at the intersection of quite disparate threads of physics, spanning decades of literature across different research communities. In this section we briefly review the ground work we need on two key topics: the flux and velocity distribution of dark matter incident upon the Earth, and the basic phenomenology of collective excitations in superfluid helium.

\subsection{Dark matter halo}
\label{sec:DMhalo}

   Estimating the rates and spectra of interactions between dark matter and Earth-bound systems requires a model of the density and velocity distribution of dark matter particles in our local neighborhood.   
Interaction rates are directly proportional to the local dark matter density, a relatively uncertain quantity (see e.g.~\cite{Read:2014qva,Benito:2019ngh,deSalas:2019pee}); we adopt here $\rho \simeq 0.4$ GeV/cm$^3$.
The velocity distribution of dark matter, $f(v) dv$ (normalized to unity), in the solar neighborhood is typically modeled as Maxwell-Boltzmann, cut off at some galactic escape velocity, $v_{esc}$; here we assume a characteristic velocity $v_0=230$~km/s~\cite{Lewin:1995rx, velocity} in the Galactic frame, with $v_{esc}=550$~km/s \cite{Smith:2006ym}.  This velocity distribution is further boosted into the rest frame of the Sun ($v_E=244$~km/s), with further modulations from the Earth's motion around the Sun neglected here~\cite{sun-velocity,flreview}.  Dark matter particles are thus incident upon terrestrial detectors at typical velocities of magnitude $v \sim 300$~km/s, but have significantly higher and lower speeds.   

  This simple form for the velocity distribution, the `standard halo model,' is a useful first approximation to the local dynamics of dark matter; its common use enables straightforward comparison between different experimental probes of dark matter.  However, the actual phase space distribution of dark matter at the Earth is poorly determined experimentally, and 
we should not expect the standard halo model to yield a precise description of the local dark matter distribution.  Firstly, numerical simulations of Milky~Way-like galaxies typically predict velocity distributions broader than Maxwellian, with more support at large speeds \cite{Kuhlen:2009vh}.  Secondly, the Milky Way is not in a steady state: smaller systems are continuously accreting onto the Milky Way, giving rise to substructure in the phase space distribution of dark matter, which can produce  localized enhancements of dark matter at relatively high velocities \cite{Lisanti:2011as,OHare:2018trr}. 

  The total elastic scattering rate evaluated at the mean velocity and the rate integrated over the full velocity distribution differ only by factors of order one.  However, the total rates above some specified threshold can be much more sensitive to the form of the velocity distribution, depending on where the threshold falls relative to the mean of the distribution.  In particular, velocity averaging is critical for determining the lowest dark matter masses that a given experiment is sensitive to, and will be a necessary component of any trade-off made between, e.g., measuring lower energy single-phonon signals versus exposing a higher-threshold detector for longer times.  For convenience we will typically quote parameters at a single representative velocity, but integrate over the standard halo model when giving total rates in Secs.~\ref{sumrulesec} and \ref{2phononsec}.

\subsection{Helium excitations}
   
\begin{figure}[h]
\includegraphics[width=6.8cm]{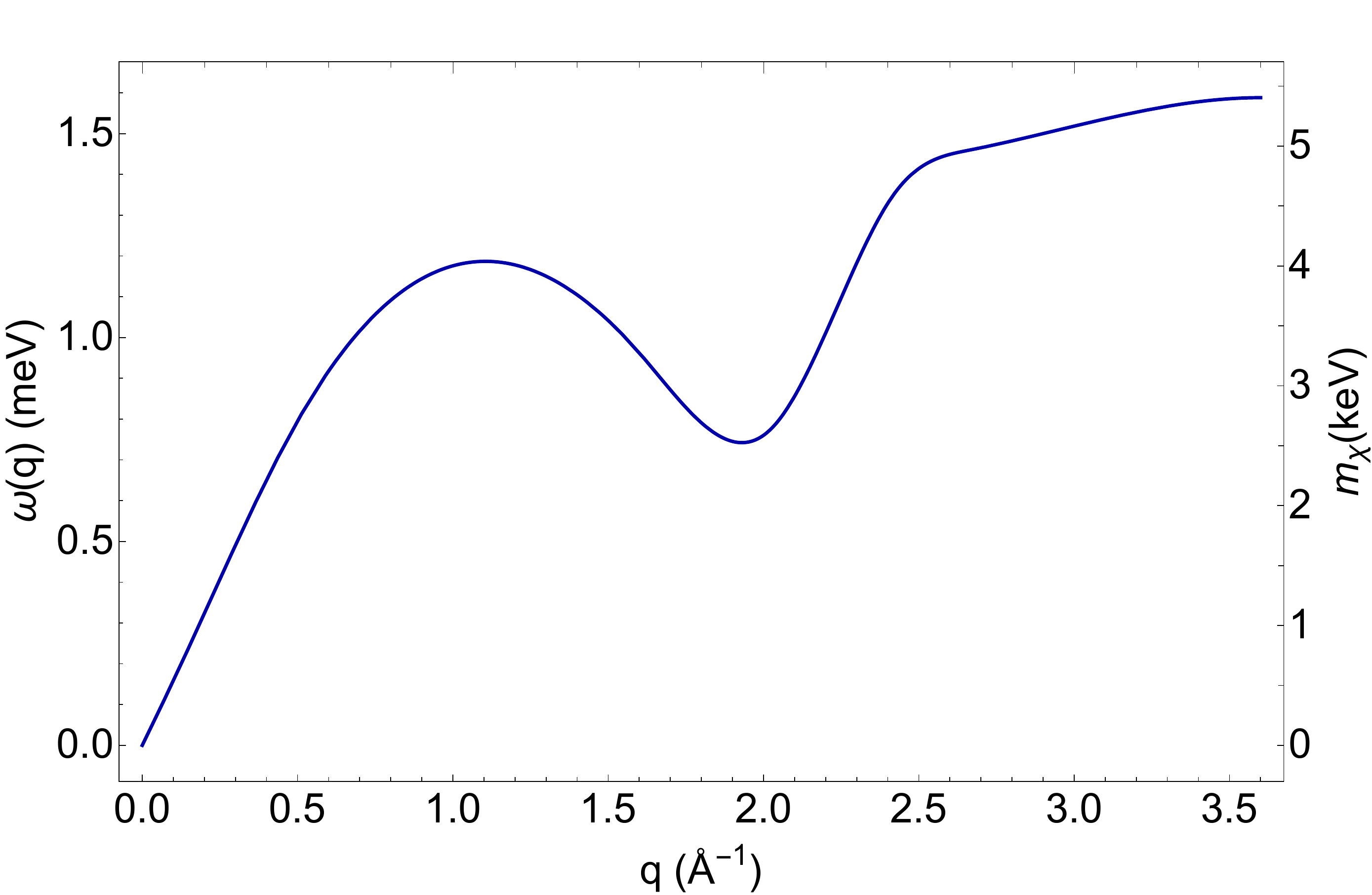}
\caption{The dispersion curve of superfluid helium~\cite{donnelly1981}.  The excitations in the range below about $q = 1$~$\rm \angstrom^{-1}$ are phonons, and in the vicinity of the $q = 1.9 $~$\rm \angstrom^{-1}$ are rotons.   Although not visible on this scale, the second derivative of the dispersion curve is positive ({\em anomalous dispersion}) below momenta $q_{\rm infl}$;  at saturated vapor pressure (SVP)  $q_{\rm infl} \approx 0.216$~$\rm \angstrom^{-1}$.   At pressure above $\sim$ 18 bar, the anomalous dispersion vanishes.   The scale on the right indicates the dark matter mass, $m_\chi$, with  kinetic energy corresponding to the scale on the left, for dark matter particles moving at (minus) the velocity, $v$, of the solar system through the galaxy. }
\label{fig:dispersion}
\end{figure}

     The de Broglie wavelength, $\lambda_{dB} =2\pi \hbar/ m_\chi v$,  of a dark matter particle of mass $m_\chi$ between $10$~keV and 1~MeV,  of order 2000 to 20{\angstrom}, is much larger than the average spacing, $\sim$ 4.5{\angstrom}, between He atoms in the superfluid.  Therefore, the $\chi$ are scattered by the helium via creation of collective modes of the superfluid -- the phonons and rotons.      The familiar phonon-roton dispersion curve is shown in Fig.~\ref{fig:dispersion} \cite{donnelly1981}.     Detection of $^4$He excitations created by a $\chi$, and measurement of their energies and directions with respect to $\vec v$ is adequate to learn the mass of the initial dark matter particle, as well as the $\chi$-$^4$He cross section.   In this paper we focus primarily on phonons.

\begin{figure}[h]
\includegraphics[width=7.51cm]{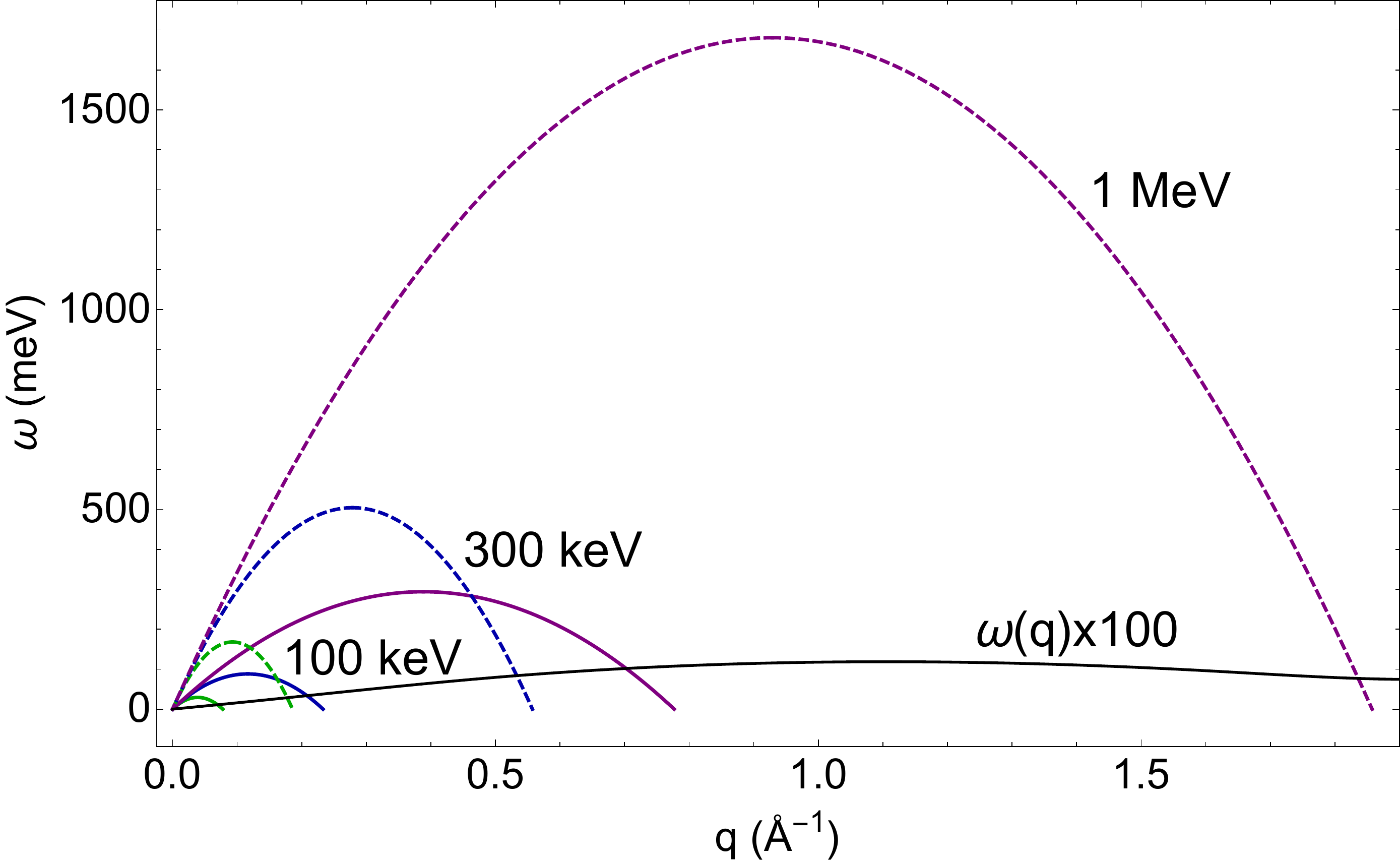}
\includegraphics[width=7.51cm]{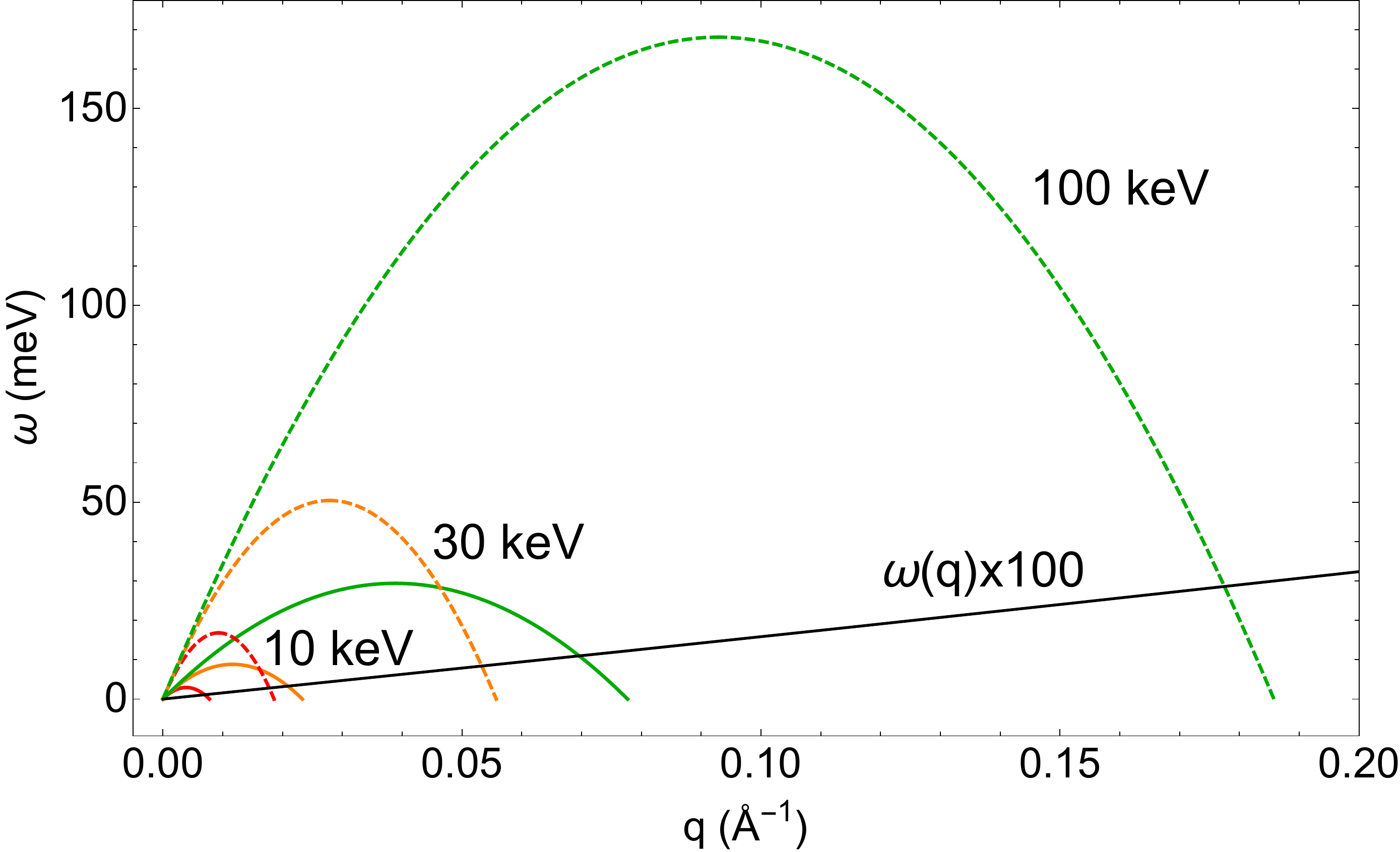}
\caption{The region of energy and momentum deposition allowed by Eq.~(\ref{omega}) for a dark matter particle of mass $m_\chi$, lies within the inverted parabolas, labelled by $m_\chi=1$ MeV, 300  keV, 100  keV (upper panel), and $m_\chi$ = 100 keV, 30 keV  and 10 keV (lower panel).  The solid lines assume velocity $v=230$ km/s, and the dashed curves $v$ = 550 km/s.   For creation of a single phonon by the dark matter particle, the energy-momentum transfer lies on the 
single phonon dispersion curve, Fig.~\ref{fig:dispersion}, shown here multiplied in energy by 100.  Only the $m_\chi$ = 1 MeV curve, for $v$ = 550, km/s extends beyond the maximum in the phonon dispersion curve.} 
 \label{fig:kinematics}  
\end{figure}

        The phonon dispersion relation, energy vs.~momentum, is approximately $\omega \simeq sq$, where $s = 2.38\times10^4$~cm/s = 1.56 meV \AA \, (at saturated vapor pressure, SVP) is the speed of sound in the superfluid.  We take $\hbar=1$ throughout.   Dark matter particles thus move at supersonic velocity, $v\sim 10^3 s$.    Importantly, between SVP and  $\sim$18 bar \cite{greywall,wyatt,sridhar} the phonon dispersion relation curves slightly upward initially,
\beq
      \omega(q) \simeq sq(1+\zeta_A q^2 + \cdots), 
    \label{anom}
\eeq
an effect known as {\em anomalous dispersion.}  At SVP,  the upward curvature stops at phonon momentum $q_{\rm infl} \approx 0.216$~$\rm \angstrom^{-1}$, where the dispersion relation has an inflection point.

   A more detailed parametrization of the phonon dispersion relation than Eq.~(\ref{anom}), valid up to $q \sim 0.9 \rm \angstrom^{-1}$,  is given by Maris \cite{maris},\footnote{Expression (2) implicitly assumes that $\omega(q)^2$ is analytic in $q^2$, and consequently that a power series expansion of $\omega(q)$ has only odd powers of $q$.   However, as pointed out by Kemoklidze and Pitaevskii \cite{KemoklidzePitaevskii}, the $r^{-6}$ falloff of the van der Waals interaction between helium atoms implies that $\omega(q)$ has even powers of $q$, beginning with $q^4$. More recently, approximations to the excitation spectrum that also contain even powers of $q$  have been made (see, e.g., \cite{Beauvois}).     The addition of such terms to the dispersion relation is not expected to alter the basic picture we develop, so to simplify the discussion we do not take them into account explicitly.}
 \beq   
    \omega(q) \simeq sq\left(1+\zeta_A q^2\left(\frac{1-q^2/q_a^2}{1+q^2/q_b^2}\right)\right),
     \label{marisparam}
\eeq   
with parameters: $\zeta_A= $1.11\AA$^{2}$, $q_a$ = 0.542\AA$^{-1}$, and $q_b$ = 0.332\AA$^{-1}$  at SVP; see also Refs.~\cite{glyde,stirling}. The parametrization (\ref{marisparam}) includes the negative curvature of the dispersion relation at higher $q$, but  does not accurately describe the peak in the dispersion relation, as shown in Fig.~{\ref{fig:dispersion}}.      At higher phonon momentum, we will use the simple parametrization \cite{khalat}   
\beq 
 \omega(q) \simeq sq(1-\zeta_N q^2 + \cdots),
 \label{normal}
\eeq 
with $\zeta_N \simeq 0.27\rm \angstrom^{2}$.

  Owing to anomalous dispersion,  single phonon final states are stable against decay only for $q$ larger than a critical value, $q_c$, which depends on the details of the dispersion relation.    For  the Maris dispersion relation (\ref{marisparam}), $q_c$ = 0.4215 \AA$^{-1}$, with $\omega(q_c)$ = 7.90 K = 0.68 meV.    (A phonon of momentum $q_c$ can decay into two collinear equal momenta phonons, $\omega(q_c)=2\omega(q_c/2)$; see details in Sec.~\ref{phdamp}.) 
Phonons produced with $q < q_c$  generate a cascade of lower momentum phonons.

\section{Dark matter scattering on superfluid helium}

   We first lay out the region of possible energy and momentum transfer, $\omega$ and $\vec q$, from a dark matter particle, $\chi$, to the helium.  The allowed energy transfer vs. momentum transfer is shown in Fig.~\ref{fig:kinematics} for representative $m_\chi$. 
For initial momentum $\vec k = m_\chi \vec v$ of the $\chi$,  the final momentum is $\vec k' = \vec k - \vec q$, and the energy transfer is
\beq
  \omega = \vec v\cdot \vec q - \frac{q^2}{2m_\chi}.
  \label{omega}
 \eeq  
 For $v \simeq$ 230 km/s, the incident momentum is $k \simeq 0.39\, m_{\rm MeV}$ \AA$^{-1}$, where $m_{\rm MeV}$ is the mass of the $\chi$ measured in MeV.    The maximum energy transfer for given $q$ occurs when $\vec q\,$ is parallel to the incident $\vec k$; then $\omega_{max}(q) = \left(kq-q^2/2\right)/m_\chi$, an inverted parabola ranging from 0 to $2k$ along the $q$ axis, 
with a maximum at $k=q$ and height $\omega_{max}(k) = k^2/2m_\chi$,  which is the maximum energy transfer from the dark matter particle.  The momentum transfer ranges from $0$ to $q_{max} = 2k$, the latter for back scattering and no energy transfer to the medium.   

   Since the energy-momentum transfers of the dark matter particle to the liquid $^4$He are so much smaller than the scales associated with the expected microscopic interactions of dark matter with the $^4$He nuclei or with the electrons, the scattering is primarily s-wave, and the interaction of a dark matter particle with a $^4$He atom can be described by a low energy pseudopotential,\footnote{If the interaction of the $\chi$ with baryonic matter is mediated by a (dark)
meson of mass $\mu$, then in lowest order the interaction is a Yukawa-like potential, $V_\mu(r) = a\mu^2e^{-\mu r}/r$, with a scattering amplitude of the form, $a/(1+(\mu q)^2)$, where $q$ is the momentum transfer from the dark matter to the helium.  The assumption of a simple pseudopotential is valid for $q\ll \mu$, in which case the scattering amplitude is simply $a$.  On the other hand if $\mu$ is smaller than the range of observable phonon momenta, $\mu \ll q$, then the scattering amplitude would behave as $a\mu^2/q^2$, increasing strongly with decreasing $q$.  If one places an upper bound on the scattering amplitude by measurements involving momentum transfers $q_0$ larger than typical phonon momenta,  then lower momentum processes can have a scattering amplitude larger by a factor $\sim (1+q_0^2/\mu^2)/ (1+q^2/\mu^2)$ than the bound.  For $q, q_0 \ll m_0$ the growth in amplitude, $\sim q_0^2/q^2$, can be significant.} 
\beq   
    V_{\chi_4}= \frac{2\pi a}{m_{r}}\delta(\vec r_\chi - \vec r_4\,),
    \label{vchi4}
\eeq 
where $a$ is the scattering length and $m_{r}$ is the dark matter--$^4$He reduced mass; for $m_\chi \ll m_4$, the $^4$He mass, $m_r \simeq m_\chi$. The total cross section for scattering of a dark matter particle on an isolated $^4$He atom is $\sigma_{\chi_4} = 4\pi a^2$.    The pseudopotential modifies the energy of a dark matter particle in liquid $^4$He  by $2\pi an_4/m_{r}$, where $n_4$ is the $^4$He equilibrium number density = 2.379 $\times 10^{22}$ cm$^{-3}$.   More complicated dependence of the dark matter particle energy on $n_4$ and the $^4$He velocity fluctuations can be ignored since the interaction of the $\chi$ with the $^4$He is weak.

  The differential rate at which dark matter particles of density $n_\chi$ deposit energy $\omega$ and momentum $q$ in the $^4$He is given in terms of the dynamical structure function of the $^4$He by
\beq
   d\Gamma = n_\chi \left(\frac{2\pi a}{m_\chi} \right)^2   2\pi n_4 S(q,\omega) \frac{d^3q}{(2\pi)^3},
 \label{gammaS}
\eeq
where $\omega$ is given by Eq.~(\ref{omega}), and the dynamical structure function is
\beq
    &&\hspace{-36pt} 2\pi n_4 S(q,\omega)\nonumber\\ &=& \int d^3 r dt \,e^{-i\vec q\cdot (\vec r\,-\vec r\,')+i\omega(t-t')}\langle \rho(\vec r,t)\rho(\vec r\,',t')\rangle \nonumber\\
   &&=\sum_f |\langle f|\rho_{-\vec q}\,|i\rangle|^2 2\pi \delta(\omega-E_f +E_i) ,
   \label{skomega}
\eeq   
with $\rho$ the $^4$He number density operator and $\rho_q = \int d^3r e^{-i\vec q\cdot \vec r}\rho(\vec r\,,t)$.  The states $i$ and $f$ are those of $^4$He in equilibrium, in the absence of dark matter, and a thermal average over states $i$ is assumed at non-zero temperature.   We assume $\omega >0$ always.

\begin{figure}[h]
\includegraphics[width=7cm]{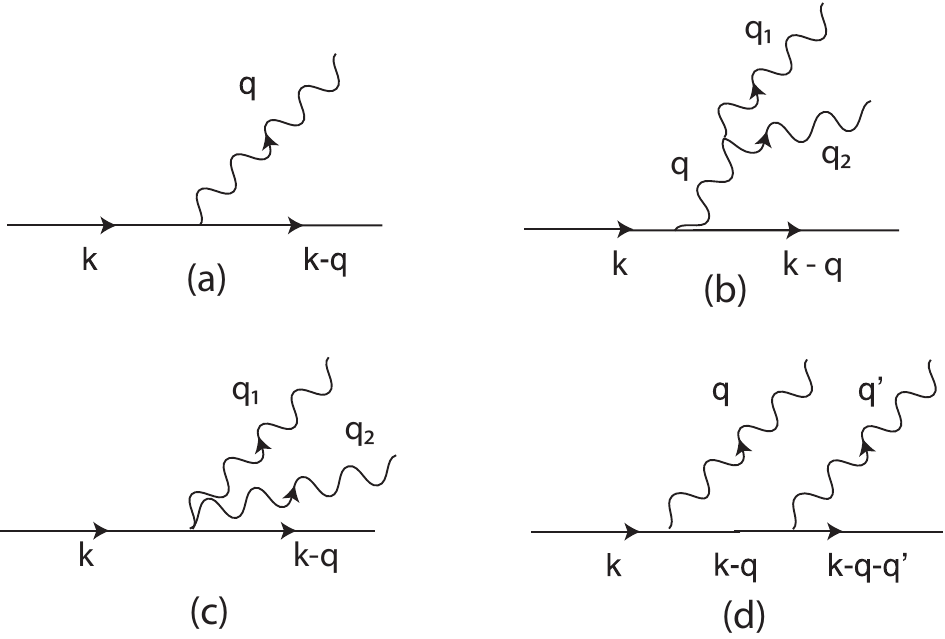}
\caption{Phonon creation by a dark matter particle.  a) single phonon creation and b) creation of a single phonon which transforms into two phonons (this process also describes direct creation of three phonons, one of which is absorbed by the dark matter particle),  c) direct creation of two phonons at the dark matter-$^4$He vertex, and d) successive creation of two phonons by the dark matter particle.    Process d) is higher order in the dark matter--helium scattering length and can be ignored in estimating the event rate for a dark matter particle creating phonons.  }
\label{fig:dmtophonons}
\end{figure}

    The Fourier transform of the $^4$He number density operator, $\rho_{-\vec q}$,
acting on a state of the liquid, can create one or more elementary excitations of the fluid of  total momentum $\vec q$, or annihilate excitations of  total momentum $-\vec q$.   Thus a dark matter particle interacting with the  $^4$He can create one or more excitations of the superfluid.   The creation of a single phonon is illustrated in 
Fig.~\ref{fig:dmtophonons}a.  This phonon can more generally transform into two or more phonons via the multiphonon interactions in $^4$He (Fig.~\ref{fig:dmtophonons}b).  In addition, a $\chi$ can directly create a pair of phonons, as shown in Fig.~\ref{fig:dmtophonons}c.   
We note that successive creations of phonons by a dark matter particle, as in Fig.~\ref{fig:dmtophonons}d, is higher order in the dark matter--helium scattering length and can be ignored.

    The rate (\ref{gammaS}) is independent of the azimuthal angle of $\vec q$ about $\vec v$.   Since the energy $\omega$ is linear in $\cos\theta_q$, where $\theta_q$ is the angle between   $\vec q$ and $\vec v$, we may replace the differential $d\cos\theta_q$ by $d\omega/qv$ and rewrite the rate as 
\beq
    \frac{d\Gamma}{dq\,d\omega} =\Gamma_0 \frac{ q}{2k^2}S(k,\omega),
\eeq
where $\Gamma_0 = \sigma_{\chi_4}\, n_\chi n_4 v$ is the rate of interaction per unit volume of the $\chi$s with a gas of non-interacting $^4$He atoms at rest.  In addition, the total rate for given momentum transfer $q$ is
\beq
   \frac{d\Gamma}{dq} =\frac{\Gamma_0}{2}   \frac{q}{k^2}\int_0^{qv - q^2/2m_\chi} d\omega\, S(q,\omega).
 \label{dgammadq}
\eeq

\section{Sum rule constraints: single phonon vs. multi-excitation emission}
\label{sumrulesec}

    Although the momentum transfers from dark matter particles are below $\sim$ 0.7 \AA$^{-1}$ for $m_\chi$ below 1 MeV and $v \sim$ 230 km/s, energy transfers for momentum transfer $q$ can be much larger than $\omega(q)$,  the energy required to generate a single phonon.  However, for low $q$ , the spectral weight of the structure function $S(q,\omega)$ is dominated by the single phonon contribution, and the most likely outcome is that for dark matter in the sub-MeV mass range a single phonon will be produced.  We can estimate the importance of multi-excitation processes from the f-sum rule (\ref{sumrule}) obeyed by $S(q,\omega)$, which at at zero temperature has the form
 \beq
   \int_0^\infty d\omega \,\omega S(q,\omega) = \frac{q^2}{2m_4}.
   \label{sumrule}
\eeq
The sum rule follows directly from the relation between $S(q,\omega)$ and the density-density correlation function in the complex frequency plane, as we recall in Appendix A; this relation also provides the basis for expanding $S(q,\omega)$ at low $q$  in terms of phonon excitations.  In Appendix B we review the structure of $S(q,\omega)$ at finite temperature.
    
     Neglecting the structure of the single phonon peak for anomalous dispersion, as described below, one can write, for $\omega>0$, 
\beq
S(q,\omega) = Z(q) \delta(\omega - \omega_q) + S_M(q,\omega),
\label{eq:sqomega}
\eeq
where $Z(q)$ is the single phonon weight, and $S_M(q,\omega)$ is the remaining multi-excitation strength. 

\begin{figure}
\includegraphics[width=8.6cm]{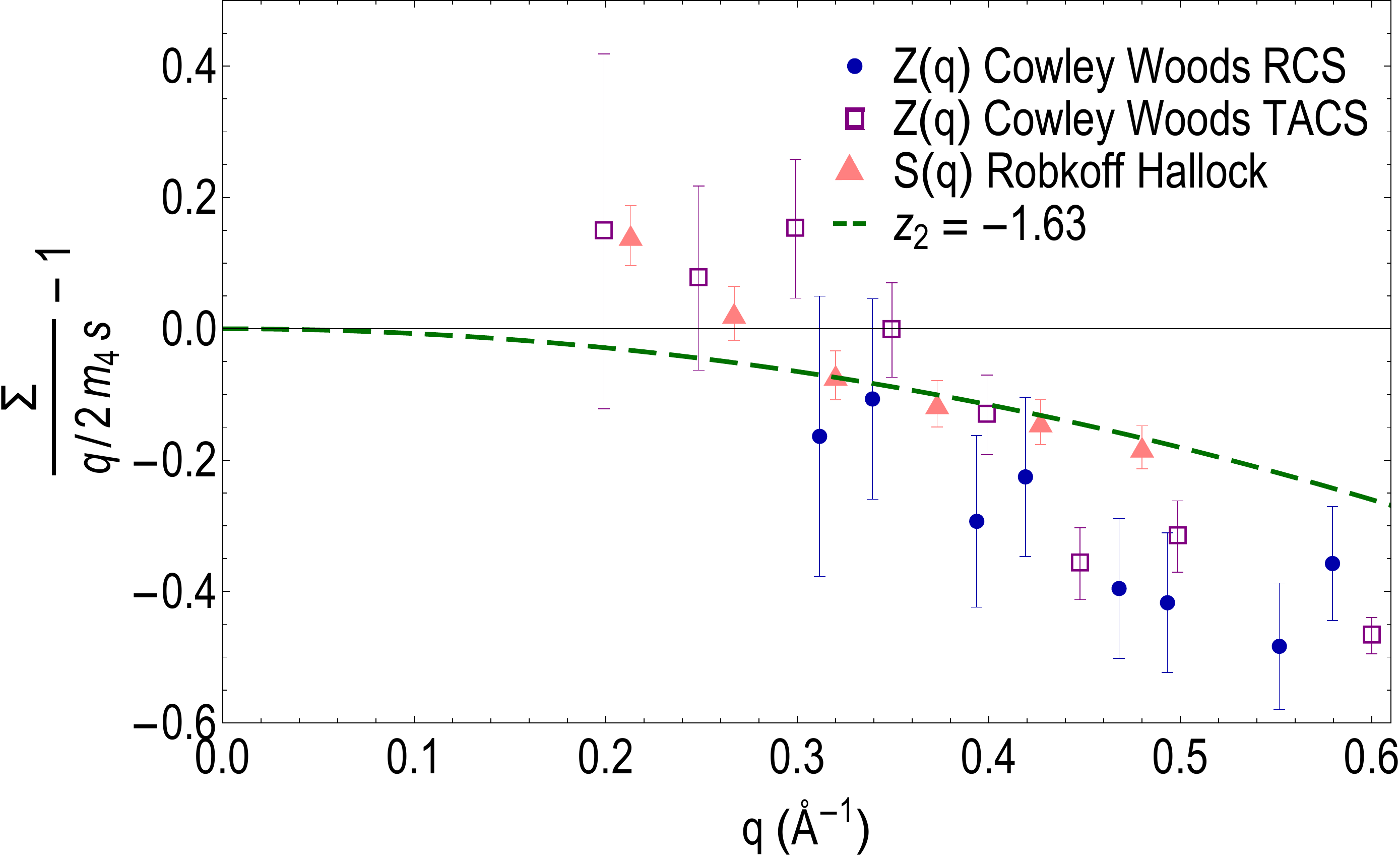}
\caption{Neutron scattering measurements of the relative deviations of the static structure function, $S(q)=\int d\omega S(q,\omega)$ and the single phonon strength $Z(q)$ from their zero temperature long wavelength single phonon value, $q/2m_4s$.   Here $\Sigma = S$ or $Z$.  
 Data points are from Robkoff and
Hallock~\cite{RobkoffHallock} at $T$ = 1.38 K ($\blacktriangle$), and from Cowley and Woods~\cite{CowleyWoods} at $T$ = 1.1 K
using the Rotating Crystal Spectrometer (RCS, $\bullet$) and the Triple Axis Crystal Spectrometer (TACS, $\square$) at Chalk River, respectively.
  The data have been corrected to zero temperature by dividing the experimental $S_T(q)$ at finite $T$ by $1+2n(\omega(q))$ and the experimental $Z(q)$ by  
$1+n(\omega(q))$, where $n(\omega(q)) = (e^{\omega(q)/T}-1)^{-1}$ is the Bose occupation factor for excitations of momentum $q$.  See discussion in Appendix B.  The dotted curve shows the sum-rule based fit,  Eq.~(\ref{zq}).}
\label{S(q)new}
\end{figure}

   The total excitation strength is bounded by the energy-weighted sum rule (\ref{sumrule}).   At zero temperature, in the absence of significant multi-excitation strength,  $Z(q) \to q^2/2m_4\omega(q)$.  With increasing $q$,
the weight of the single phonon peak in Eq.~(\ref{eq:sqomega}) is reduced from $q^2/2m_4\omega(q)$ by direct creation of two (or more) phonons and rotons;  the sum-rule arguments given in Refs.~\cite{pines-woo,liu-woo} and reviewed in Ref.~\cite{sridhar} indicate that
\beq
  Z(q) =\frac{q^2}{2m_4\omega(q)} \left(1-z_2\left(\frac{q}{m_4s}\right)^2+\dots\right),
  \label{zq}
\eeq   
where $z_2 \simeq 1.63$, a value consistent with neutron scattering experiments \cite{RobkoffHallock,CowleyWoods,Svensson}; see Eq.~(\ref{expansions}) for the expansion to higher order in $q$.  The correction $\sim z_2$ is shown as the dashed curve in Fig.~\ref{S(q)new}.

   The f-sum rule together with Eq.~(\ref{zq}) then implies that for small $q$,
\beq
   \int_0^\infty d\omega\, \omega S_M(q,\omega) =z_2 \left(\frac{q}{m_4s}\right)^2 \frac{q^2}{2m_4},
   \label{smsum}
\eeq
plus terms of relative order $q^6$.   Thus multi-excitations contribute a fraction 
\beq
   \int_0^\infty d\omega\, \omega S_M(q,\omega)\Big/\int_0^\infty d\omega\, \omega S(q,\omega) = z_2 \left(\frac{q}{m_4s}\right)^2
   \label{sumruleM}
\eeq
to the sum rule at small $q$.  At $q\simle  0.35$\AA$^{-1}$, with $m_4s$ = 
1.50\,\AA$^{-1}$, the multi-excitation contribution is $\simle$ 10\% of the single phonon contribution.  

  Similarly, the static structure function is
\beq
 S(q) = \int d\omega S(q,\omega) = Z(q) + S_M(q),
\eeq 
where the multi-excitation contribution is
\beq
   S_M(q) = \int_0^\infty d\omega  S_M(q,\omega).
\eeq
Since the multi-excitation weight is at $\omega \ge sq$, Eq.~(\ref{smsum}) implies that for small $q$, 
\beq
  S_M(q) \le  \int_0^\infty  d\omega\,  \frac{\omega}{sq} S_M(q,\omega) = \frac{z_2}{2}\left(\frac{q}{m_4s}\right)^3,
 \label{bound}
\eeq
or equivalently,
\beq
  \frac{S_M(q)}{S(q)} \le z_2 \left(\frac{q}{m_4s}\right)^2,
\eeq
the same fraction as in the sum rule, Eq.~(\ref{sumruleM}).

    The multi-excitation rate, $d\Gamma_M/dq$, for given $q$, compared to the single phonon rate, is similarly bounded.   We write
\beq
 \frac{d\Gamma}{dq} =\frac{d\Gamma_1}{dq} + \frac{d\Gamma_M}{dq},
\label{3gammas}
\eeq
in terms of the single phonon contribution  
\beq
  \frac{d\Gamma_1}{dq} = \frac{\Gamma_0}{2}\frac{q}{k^2} Z(q) \simeq  \frac{\Gamma_0}{4m_4s}\frac{q^2}{k^2},
   \label{onephonon}
\eeq
where $q\le 2k$,
and the multi-excitation contribution
\beq
  \frac{d\Gamma_M}{dq} = \frac{\Gamma_0}{2}\frac{q}{k^2} \int_0^{qv - q^2/2m_\chi} d\omega\, S_M(q,\omega).
\eeq
Since the integral is bounded above by $S_M(q)$, we find from Eq.~(\ref{bound}), the bound on the multi-excitation rate, 
\beq
   \frac{d\Gamma_M}{dq} \le \frac{\Gamma_0}{4k^2} \frac{z_2q^4}{(m_4s)^3}.
\eeq
Comparing with Eq.~(\ref{onephonon}), we have
\beq
    \frac{d\Gamma_M}{dq}\Big/\frac{d\Gamma_1}{dq} \le z_2\left(\frac{q}{m_4s}\right)^2,
    \label{sumrulerates}
\eeq
the same ratio as the contributions to the sum rule.   The multi-excitation rate at $q\simle 0.4$\AA$^{-1}$ is similarly $\simle$ 10\% of the single phonon contribution.   The sum-rule argument implicitly takes into account  the momentum dependence of the matrix elements for producing multiphonon states as well as the available phase space.   We emphasize that this bound does not depend on whether the dispersion is anomalous or not.

  When the phonon dispersion is normal, single (on shell) phonons cannot decay into two or more phonons.  However with anomalous dispersion 
a single phonon can decay into two, and there is no longer a clean distinction in $S(q,\omega)$ between damped single phonons and multi-phonon states; the effect, as we see below, is to spread the single phonon peak.   Off-shell phonons can decay
into two phonons if $\omega-\omega(q)>0$ for normal dispersion and if 
$\omega >2\omega(q/2)$, or equivalently $\omega
-\omega(q)>-(\omega(q)-2\omega(q/2))$, for normal dispersion.  For 
$\omega -\omega(q)\gg|\omega(q)-2\omega(q/2)|$ the decay rate becomes
independent of the sign of the dispersion.

    We turn to estimating the rate of single phonon events.  For simplicity we assume here that the single phonon spectrum cuts off at $q$ of order the  $^4$He Debye wavevector $k_D = 1.089 \,{\rm \AA}^{-1}$, defined in terms of the $^4$He number density by $n_4 = k_D^3/6\pi^2$.  
Equation~(\ref{onephonon}) implies that the integrated one-phonon event rate, for a detector with a lower energy threshold  $\omega_0$ with $\omega_0/s \le 2k = 2m_\chi v \le k_D$, is
\beq
   \hspace{-12pt} \Gamma_1 (\omega >\omega_0) &=& \int_{\omega_0/s}^{2k} dq \frac{\Gamma_0}{m_4 s} \frac{q^2} {4k^2} \nonumber\\
&=&   \frac{2n_4\rho_\chi\sigma_{\chi_4}}{3m_4}  \frac{v^2}{s} \left(1-\left(\frac{\omega_0}{2sm_\chi v}\right)^3\right), %
\label{gammaone}
\eeq
while in the limit of large $m_\chi$, with $\omega_0 \le k_D \le 2k$, one has
\beq
    \hspace{-12pt} \Gamma_1 (\omega >\omega_0) &=& \int_{\omega_0/s}^{k_D} dq \frac{\Gamma_0}{m_4 s} \frac{q^2} {4k^2} \nonumber\\
&=& \frac{n_4\rho_\chi\sigma_{\chi_4}}{12m_4 s}  \frac{k_D^3}{m_\chi^3 v} \left(1-\left(\frac{\omega_0}{sk_D}\right)^3\right).%
\label{gammaone2}
\eeq

   The prefactor of $\left(1-\left(\omega_0/2sm_\chi v\right)^3\right)$ in Eq.~(\ref{gammaone}) is
\beq  
  \simeq  3.0 \times 10^{-9}\left(\frac{\sigma_{\chi_4}}{10^{-40}{\rm cm^2}}\right)\left(\frac{v}{230 {\rm km/s}}\right)^2\,{\rm s^{-1}cm^{-3}},
   \eeq
where $\rho_\chi = n_\chi m_\chi \simeq 0.4$ GeV/cm$^3$ is the local dark matter density.   The integrated rate for small $m_{\chi} v$ is proportional to $\rho_x$, and depends on $m_\chi$ only through the factor $\omega_0/2sm_\chi v$.     
The factor $2sm_\chi v$ is $\sim$ 0.12 ($m_\chi$/100~keV)($v$/230~km/sec)~meV.  

\begin{figure}[h]
\includegraphics[width=8.5cm]{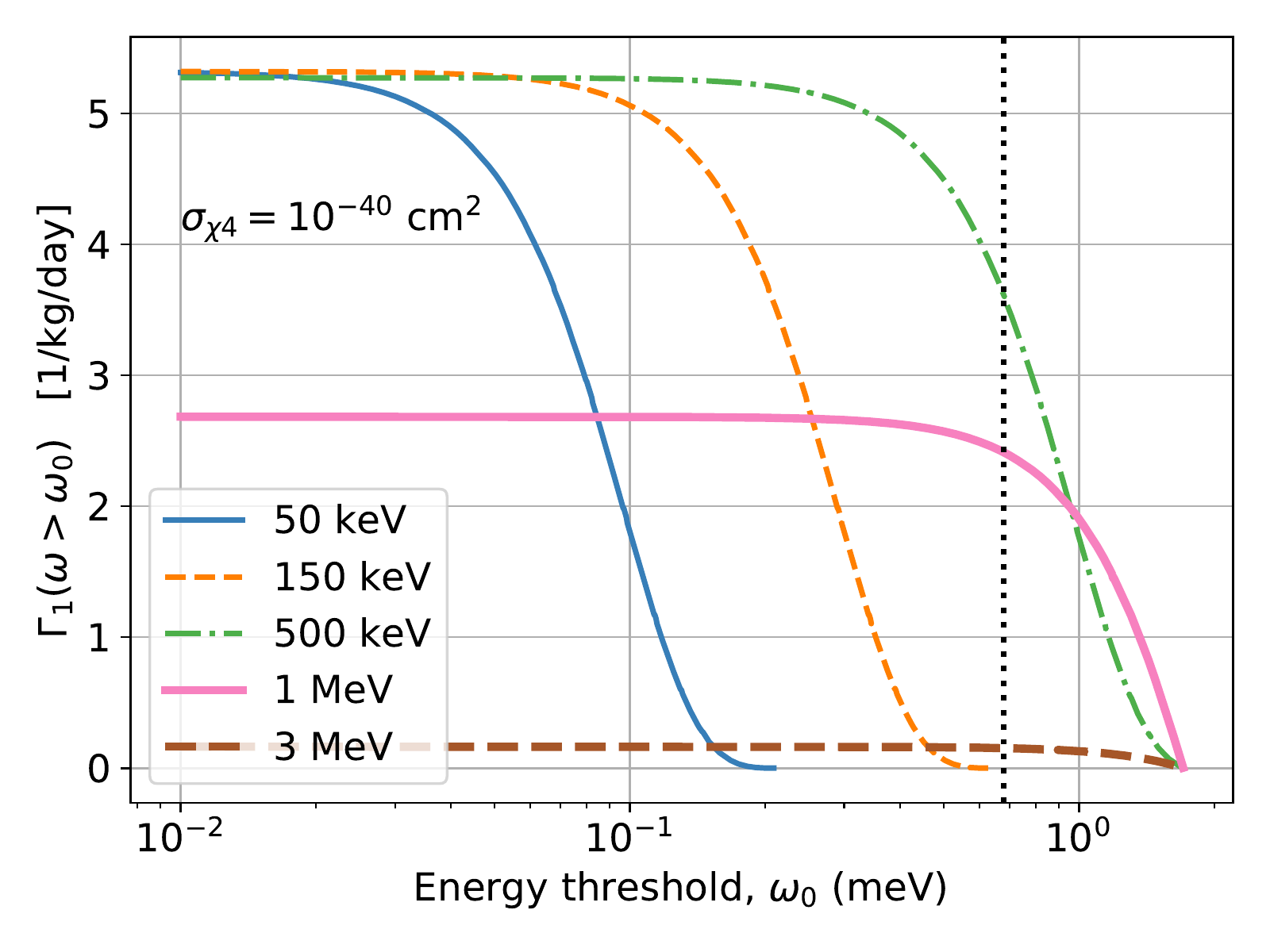}
\caption{Rate of single-phonon events per kg of $^4$He for $\sigma_{\chi 4}=10^{-40}$~cm$^2$ and a selection of masses $m_{\chi}$, as a function of limiting energy threshold $\omega_0$.  Rates are integrals of  $\Gamma_1 (\omega >\omega_0)$ (Eqs.~(\ref{gammaone}) and (\ref{gammaone2})) over the standard halo model, following the parameters in Sec.~\ref{sec:DMhalo}. The vertical dotted line indicates $\omega(q_c)$ at 1~bar, below which phonons can split into two or more phonons.  The rapid falloff for MeV-scale $m_\chi$ arises from the lack of phase space for single phonon production.}
\label{fig:threshplot}
\end{figure}

   Figure~\ref{fig:threshplot} shows the one-phonon rate integrated over the standard halo model described above, $\int dv f(v) \Gamma_1 (\omega >\omega_0)$,
for various $m_\chi$.  For low $m_{\chi}$ the zero-threshold rate scales with the mean square velocity 
$
  \int dv \, v^2 f(v) =(369 \,{\rm km/s})^2,
$
and the single phonon rate is given by
\beq
  \langle \Gamma_1(\omega>0)\rangle \sim 8.9 \times 10^{-9}{\rm cm}^{-3}{\rm s}^{-1} (\sigma_{\chi_4}/10^{-40}{\rm cm}^2).
  \label{onephononnum}
\eeq 
Note that the phase space for creating single phonons falls rapidly for $m_{\chi} \gtrsim 500$~keV, as seen in Fig.~\ref{fig:threshplot}.

   More generally, the contribution of the single phonon peak to the interaction rate of a dark matter particle with the helium (with $Z(q)\to q^2/2m_4\omega(q)$) is
\beq
    \frac{d\Gamma_1}{dq\, d\cos\theta_q} = \frac{\Gamma_0}{4m_\chi^2 v} \frac{q^4}{m_4\omega(q)}\delta(\omega-\omega(q)).
    \eeq  
Integrating over $q$ using Eq.~(\ref{omega}), we find the angular distribution,
\beq
   \frac{d\Gamma_1}{d\cos\theta_q} \simeq \frac{2m_\chi\Gamma_0}{vm_4s}(v\cos\theta_q-s)^2,
   \label{eq:1phononRate}
\eeq
with the restriction that $0< v\cos\theta_q-s \simle k_D/2m_\chi$.

\begin{figure}[h]
\includegraphics[width=8cm]{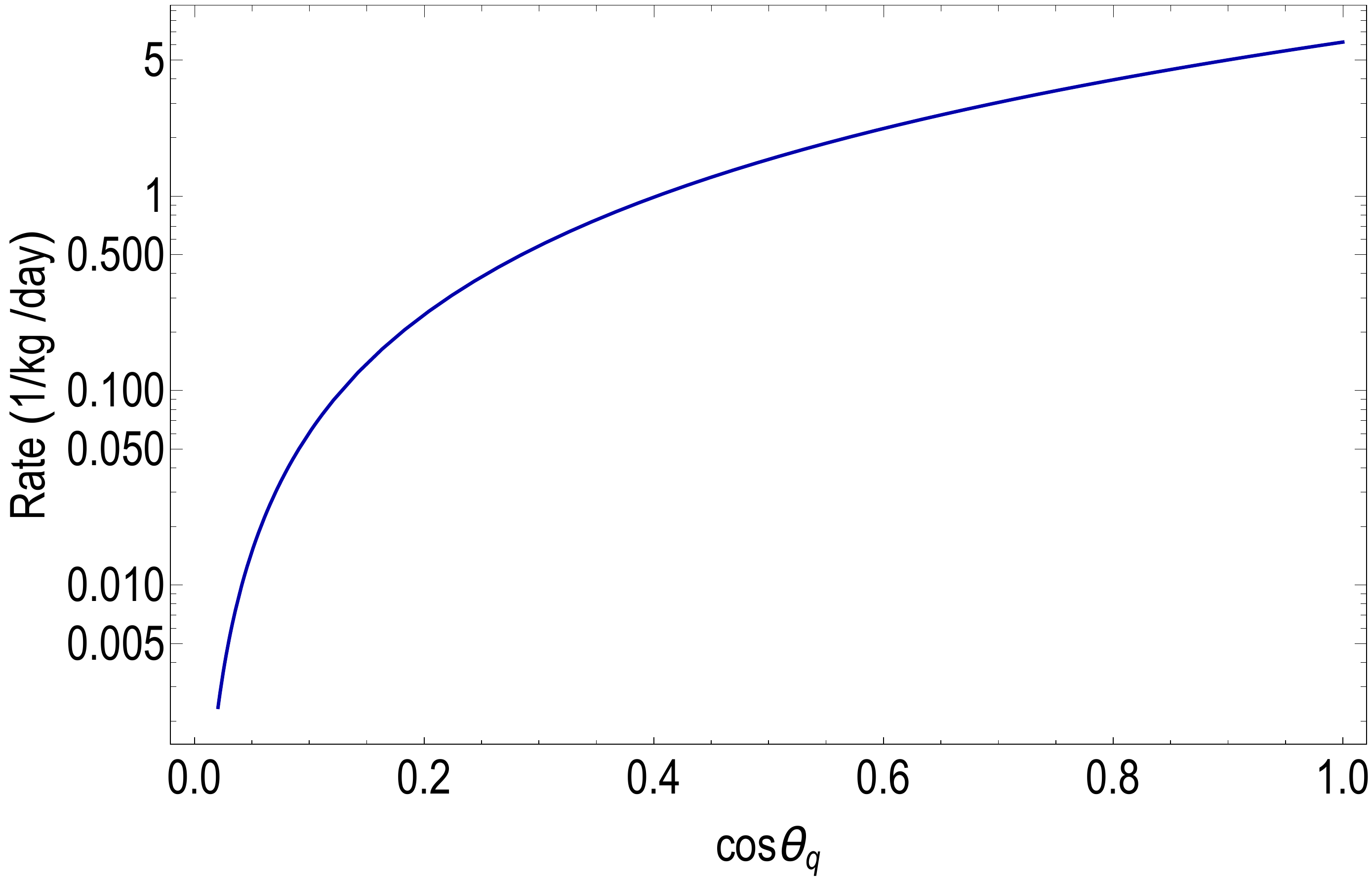}
\caption{Event rate for dark matter particles producing single phonons in the normal dispersion region, in units of events/kg/day.  The dark matter--helium cross section is taken to be $\sigma_{\chi_4} = 10^{-40}$ cm$^2$, $v$ = 230 km/s,  and the detector acceptance to be $\Delta \cos\theta_q = 0.1$.   Kinematics requires $v\cos\theta_q$ to lie in the interval $s$ to $s+ 
 k_D/2m_\chi$.}
\label{fig:1phononRates}
\end{figure}

  Figure~\ref{fig:1phononRates} shows the one-phonon rate, Eq.~(\ref{eq:1phononRate}), as a function of $\cos\theta_q$, assuming a cross-section, $\sigma_{\chi_4} = 10^{-40}$ cm$^2$, and a nominal velocity $v$ = 230 km/s, and $2m_\chi v\simle k_D$.
  
   Measurement of the momentum and energy transfers $\vec q$ and $\omega$ from detection of either single phonon or phonon pair production, implies that the mass of the dark matter particle is
\beq  
   m_\chi = \frac{q^2}{2(qv\cos\theta_q - \omega)}.
\eeq 
For single phonon detection  $\omega \simeq sq$, and thus
\beq  
   m_\chi = \frac{q}{2(v\cos\theta_q - s)}.
\eeq 
Except for emission at $\theta_q\simeq \pi/2$, one can neglect the $s$ and conclude that $m_\chi \simeq 1.36\, q /\cos\theta_q$, where $m_\chi$ is measured in MeV and $q$ in $\angstrom^{-1}$.

 \section{Phonon damping}
 \label{phdamp}

       The principal damping mechanism of a phonon, either on- or off-shell, is decay into two phonons when kinematically allowed.  A single phonon produced on-shell will decay into two phonons as long as its momentum is less than the critical $q_c$ (= 0.4215 \AA$^{-1}$ for the dispersion relation (\ref{marisparam})).   If $q>q_c$  the phonon is stable.   However, production of two phonons via the process in Fig.~\ref{fig:dmtophonons}b remains possible for all $q$ if the single phonon is sufficiently off-shell. 

    To understand the splitting of an on-shell phonon of momentum $q$ into a pair of phonons of momenta $q_1$ and $q_2$ (where in this discussion we assume $q_1\le q_2$), we note that the dispersion curve (\ref{marisparam}) has a number of critical points.  Beyond the  inflection point, where $\omega''(q_{\rm infl})=0$, at $q_{\rm infl}$ = 0.216 \AA$^{-1}$, the  slope of the dispersion curve equals the zero momentum sound speed $s$ at $q_s$ = 0.377 \AA$^{-1}$;  up to $q_s$ a phonon of momentum $q$ can turn into a pair of phonons, where the smaller of the momenta $q_1$ can be  arbitrarily small.   The critical momentum
$q_c$ = 0.422 \AA$^{-1}$ is where  $\omega(q_c)=2\omega(q_c/2)$, so that one on-shell phonon can decay into two collinear equal momenta phonons.   Phonons of momentum beyond $q_c$ can no longer split into pairs of phonons.
Between $q_s$ and $q_c$ phonons can still turn into pairs, but with both $q_1$ and $q_2$ finite.   
Up to $q_c$,   $q_1$ takes on its maximum possible value when $q_1=q_2 \gtrsim q/2$, with equality at $q=q_c$.    Lastly we note that $\omega(q) = sq$ again at $q= q_a$ = 0.542 \AA$^{-1}$.   .
 
\begin{figure}
\begin{minipage}{0.95\hsize}
\includegraphics[width = 0.9\textwidth]{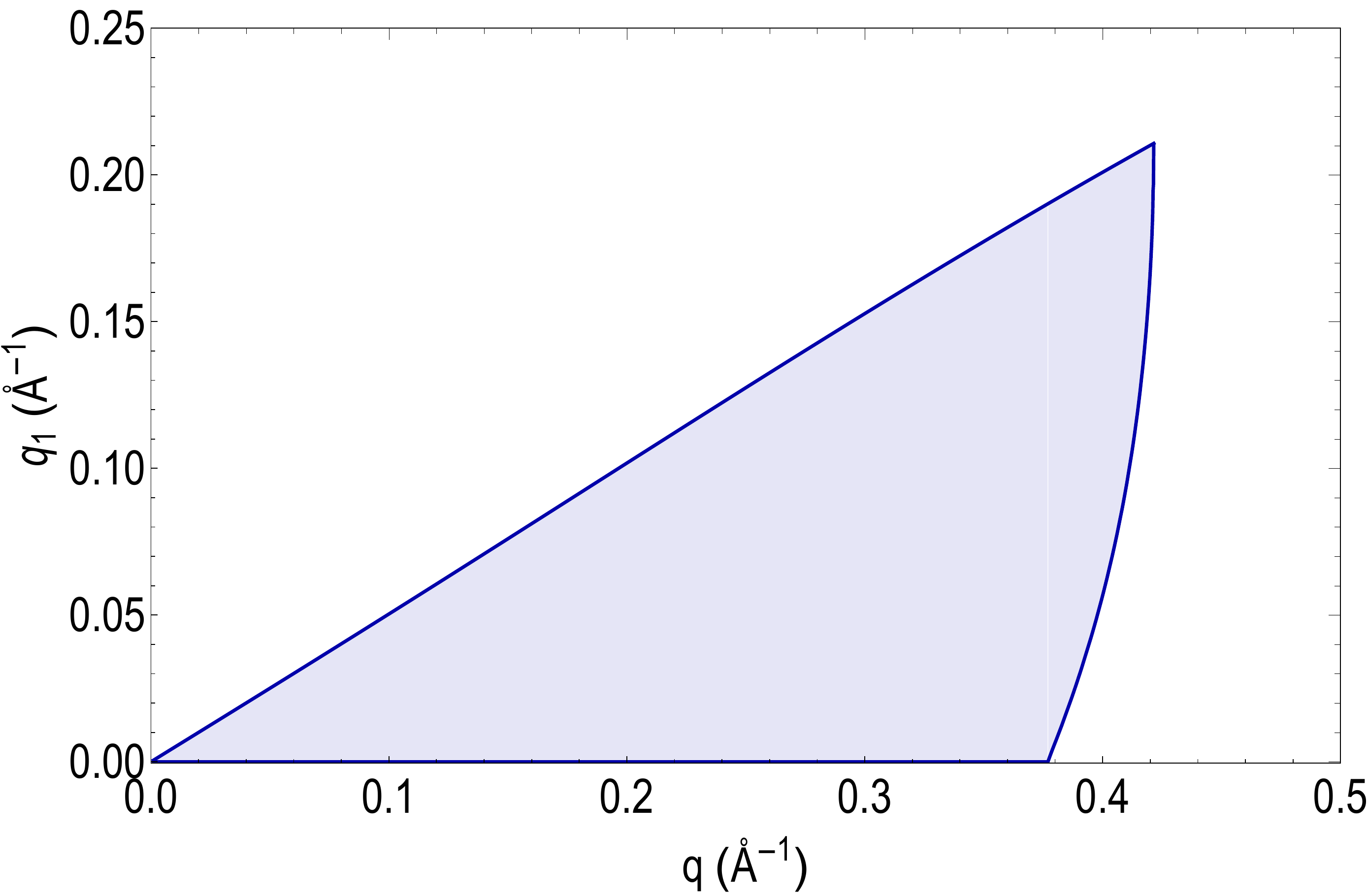}
\end{minipage}
\begin{minipage}{0.95\hsize}
\hspace{-8pt}\includegraphics[width = 0.95\textwidth]{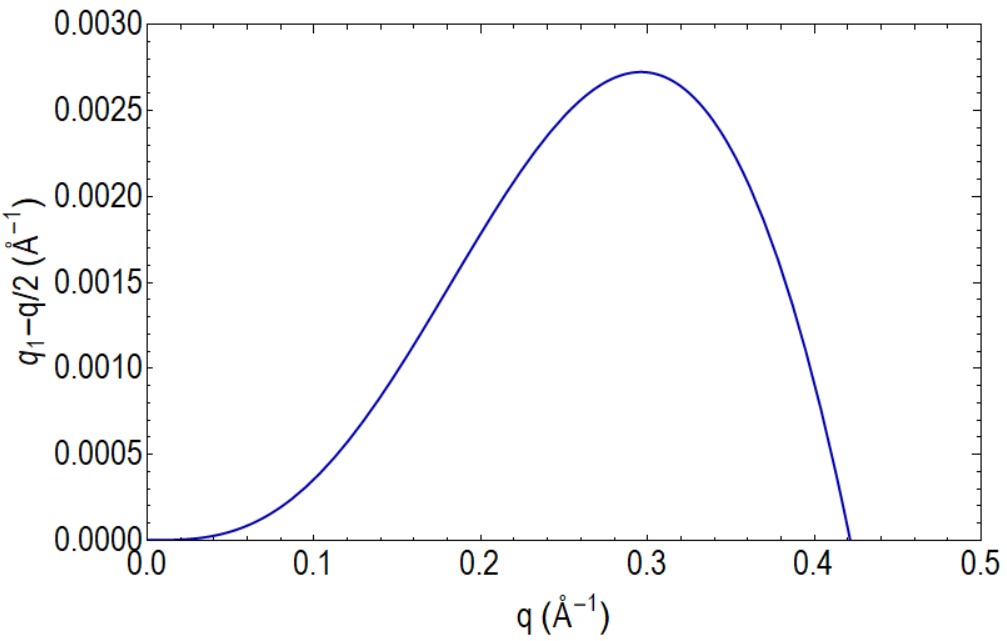}
\end{minipage}
\caption{The range of $q_1$, the smaller of $q_1$ and $q_2$ in the single splitting of an on-shell phonon of momentum $q$ into phonons $\vec q_1$ and $\vec q_2$, as a function of $q$, is shown as the shaded region The lower panel shows the fine structure of the upper limit $q_1$ compared with $q/2$, on a scale $\sim 0.01$ of the that in the upper panel.}
\label{fig:q1range}
\end{figure}

  In the decay of a phonon of momentum $q$ into $q_1$ and $q_2$, (again with $q_1 \le q_2$, so that one does not have symmetry under $q_1\leftrightarrow q_2$), the phonons are collinear only if $q_1=q_{1,\rm min}$.
For $q_1>q_{1,\rm min}$ the two phonons are at a finite angle with respect to $\vec q$.   At $q_1=q_{1,\rm max}$, one has $q_1=q_2$.
 The range of $q_1$ in the splitting of an on-shell phonon of momentum $q$, as a function of $q$, is shown in the upper panel of Fig.~\ref{fig:q1range}.   Owing to the smallness of the anomalous dispersion, the upper limit is only slightly above $q/2$, as shown in the lower panel of Fig.~\ref{fig:q1range}.

     We now examine the rate of decay $\gamma_2(q,\omega)$ of a phonon of momentum $q$ and (possibly off-shell) energy $\omega \ge \omega(q)$ into two phonons.  In a system of finite volume $\Omega$ (which we take to infinity in the end) the matrix element for a phonon of momentum $\vec q$ to
generate a pair of phonons $\vec q_1$ and $\vec q_2$ is  $\langle \vec q_1\,, \vec q_2\,|V|\vec q\,\rangle \delta_{\vec q_1+\vec q_2\, ,\vec q\,} /\sqrt\Omega$  \cite{ebner}, where 
\beq  
  \langle \vec q_1\,, \vec q_2\,|V|\vec q\,\rangle  &=&  \sqrt \frac{q^2q_1^2q_2^2s^4}{8m_4n_4\omega(q)\omega(q_1)\omega(q_2)} \times \nonumber \\
    &&\times
     \left(2u-1+\hat q_1\cdot\hat q_2 + (\hat q_1+\hat q_2)\cdot \hat q)\right), \nonumber\\
    \label{V3}
\eeq
with $u = (n_4/s)\partial s/\partial n_4$  the phonon Gr\"uneisen  parameter, $\approx$ 2.843 at SVP.     The latter angular term  is bounded above by $2(u+1)$ and below by $2(u-1)$.  In the following, rather than going through a complicated calculation involving the phonon angles  we replace this factor by $2(u+\nu)$, where for back-to-back phonons $\nu=-1$ and for collinear phonons, $\nu=1$.  The rate at which a phonon of momentum $q$ and energy $\omega$ decays into two phonons of momenta $\vec q_1$ and $\vec q_2$ is
\beq
 \label{gamma}
   \gamma_2(q,\omega) &=& \frac12
    \int\frac{d^3 q_1}{(2\pi)^3}  \frac{d^3 q_2}{(2\pi)^3}  |\langle \vec q_1\,, \vec q_2\,|V|\vec q\,\rangle|^2
   \\ &&\times(2\pi)^4\delta(\omega-\omega(q_1)-\omega(q_2)) \delta(\vec q - \vec q_1 -\vec q_2).  \nonumber
   \label{gammertwo}
\eeq
The factor 1/2 compensates for double counting the two-phonon states (since $|\vec q_1 \vec q_2\rangle = |\vec q_2 \vec q_1\rangle$).
In the calculation below the curvature of the phonon spectrum is important in the energy conserving delta function, but not in the matrix elements; thus we replace the various $\omega(q_i)$, in the prefactors in  Eq.~(\ref{V3}) by their values for linear dispersion, $sq_i$.  
   
   To evaluate the integral in Eq.~(\ref{gamma}), we use momentum conservation to do the $q_2$ integral, so that $\vec q_2 \to\vec q-\vec q_1$, and find
\beq
   \gamma_2(q,\omega)& =&\frac{(u+\nu)^2}{m_4n_4} \int \frac{dq_1}{8\pi}d\cos\delta_1\, q_1^3 q_2 \times \nonumber\\ &&\hspace{48pt}\times \delta(\omega-\omega(q_1)-\omega(q_2)), 
\eeq
where $\delta_1$ is the angle between $\vec q$ and $\vec q_1$.  The relation $q_2^2 = q^2 + q_1^2 -2qq_1\cos\delta_1$ implies that
\beq
    \int d\cos \delta_1 \,\delta(\omega-\omega(q_1)-\omega(q_2)) = \frac{q_2}{qq_1 d\omega(q_2)/dq_2},
    \label{intcosdelta1}
\eeq
where $\Theta$ is unity if there is an angle $\delta_1$ for which the argument of the energy delta function vanishes, and zero otherwise;
hence
\beq
   \gamma_2(q,\omega)& =&  \frac{(u+\nu)^2 s}{8\pi m_4n_4} \int dq_1   \frac{q_1^2q_2^2 }{\partial \omega_2/\partial q_2}
   \label{gamma-e} \\
   &=& \frac{(u+\nu)^2 s\omega}{8\pi m_4n_4} \int_0^1 dx   \frac{q_1^2q_2^2 \Theta }{d \omega_1/d q_1\,\,d \omega_2/d q_2}. \nonumber\\
  \label{gamma-d}
\eeq
In the lower equation we integrate with respect to $x \equiv \omega_1/\omega$ at fixed $\omega$.  Note that $\omega_2/\omega = 1-x$. 
The range of $q_1$ is determined by the requirement that there be an angle $\delta_1$ for which the argument of the energy delta function vanishes, and $\Theta$ includes the same restrictions in terms of $x$.  

     Were the dispersion normal, the decay of one phonon to two could only take place for $\omega>\omega(q)$.  
We explicitly evaluate the integral in Eq.~(\ref{gamma-e}) for small $q$ and $\omega-\omega(q) \ll sq$, using the dispersion relation in the form (\ref{normal}).   For given $\omega-\omega(q)$, the minimum $q_1$ is achieved for $\vec q_1$ anti-parallel to $\vec q$, and from energy conservation,  $q_1^{min} \approx (\omega-\omega(q))/2s$.   On the other hand, the maximum $q_1$ is achieved for $\vec q_1$ parallel to $\vec q$; again from energy conservation,  $q_1^{max} \approx 3(\omega-\omega(q))/s\zeta_N q^2$.  For small $q$ in both limits, $q_2\approx q$, and the integral in Eq.~(\ref{gamma-e}) is approximately $(q_1^{max})^3/3s$.  Altogether then
\beq
   \gamma_2(q,\omega) = (u+1)^2\frac{\pi q^2}{2m_4}\left(\frac{\omega-\omega(q)}{sk_D\kappa_N}\right)^3 \equiv {\cal C}(\omega-\omega(q))^3,  \nonumber\\
 \label{gammau}
\eeq  
for $\omega\simge\omega(q)$;
here $\kappa_N \equiv 3\zeta_N q^2$;  at SVP, $\kappa_N = 0.98 (q/k_D)^2$.
In deriving (\ref{gammau}) we have 
replaced $\nu$ by +1 since for small $q$ the integral is dominated by collinear phonons.  The damping vanishes at $\omega = \omega(q)$, as expected. 
For $\omega$ sufficiently close to $\omega(q)$, $S$ [from Eq.~(\ref{sgamma}) with  $\cal R$ neglected] reduces to a delta function 
\beq    
  S(q,\omega)  \to \frac{q^2}{2m_4\omega(q)}\delta(\omega - \omega(q)),
  \label{sdelta}
\eeq
plus a continuum, corresponding to multiphonon excitations, which, as is schematically illustrated in Fig.~\ref{Sscartoon}, goes to zero as $\omega-\omega(q)$ when $\omega \to \omega(q)$ from above (see Eq.~{\ref{snorm} below.

   On the other hand the direct two phonon contribution to $S_2(q,\omega)$, determined by the matrix element (\ref{rho2}) below, has the same structure in $\omega$ as $\gamma_2(q,\omega)$, and is of order $(\omega - \omega(q))^3$.   The delta function, Eq.~(\ref{sdelta}), corresponds to a sharp single phonon line, which exhausts the f-sum rule (\ref{sumrule}).
In the f-sum rule, the integral over the multiphonon background is cancelled by renormalization corrections to the delta function from the real part of the phonon self-energy, $\cal R$, neither of which we consider here.

  With a purely linear spectrum, decay of one phonon into two is allowed for all energies greater than $sq$ and forbidden for energies less than $sq$, so $\gamma_2(\omega, q)$ is discontinuous at $\omega=sq$.      
  
      For anomalous dispersion with $q\le q_s$, all low momentum on-shell phonons can decay, and the range  of the $x$ integral in Eq.~(\ref{gamma-d}) is 0 to 1.    For small $q$, and $0\le \omega-\omega(q) \ll sq$ we find 
    \beq  
      \gamma_2(q,\omega) = (u+1)^2 \frac{ \pi q^5}{40 m_4k_D^3} \equiv \frac{q^5s}{\eta}.
         \label{xs}
\eeq
As $\omega$ decreases below $ \omega(q)$, decay is still possible, with $\gamma_2$ vanishing for $\omega \le 2\omega(q/2)$, as illustrated in Fig.~\ref{Sscartoon}.

  While in this calculation we assumed  anomalous dispersion in the simplified form (\ref{anom}),  this result is independent of the details of the anomalous dispersion, as long as it is small, and holds equally well for the more accurate dispersion relation (\ref{marisparam}). 

  The corresponding phonon mean free path as limited by decay into two phonons is    
\beq
    \ell(q) = \frac{s}{\gamma_2}= \frac{\eta}{q^5} = \frac{1.67 {\rm\AA}}{q^5},
    \label{mfp}
\eeq
with $q$ measured in \AA$^{-1}$ in the final expression.\footnote{Similarly the rate of absorption of a phonon on a thermal phonon (the Landau process of two phonons to one) is given by (\ref{xs}), only with $q^5$ replaced by $q(2\pi T/s)^4$ for $q\ll T/s$ \cite{spartak,pierre,LandP} .   For phonon energies large compared with those of thermal phonons, the absorption rate varies as $q^2(T/s)^3$ and this process can be ignored.}$^,$\footnote{ 
This mean free path is the length for a single phonon of momentum $\vec q$ in liquid helium to decay into two almost collinear phonons when no other excitations are present initially.  When many phonons are present,  this mean free path determines the rate at which phonons moving in approximately the same direction come into thermal equilibrium with each other due to one phonon decaying into two, and the inverse process in which two almost collinear phonons create a single phonon.  The characteristic mean free paths for thermal conductivity or viscosity \cite{greywall,jltp} are very much larger because processes involving nearly collinear phonons are ineffective in degrading heat currents or stresses: to dampen these disturbances requires establishing equilibrium between phonons moving in directions differing by angles $\sim 1$ radian, and this can only be done by a sequence of decays and coalescences that equilibrate phonons moving in slightly different directions \cite{maris, Benin}.  The difference between $\ell$ and the mean free path for thermal conduction or viscosity is similar to the difference between the total cross section for scattering of a particle and the transport cross section in the kinetic theory of gases and the theory of impurity resistivity in metals \cite{LandLX}.
} 

   Near $\omega=\omega(q)$, $S(q,\omega)$ for anomalous dispersion is essentially Lorentzian;  the spread in the peak at $\omega= \omega(q)$ is found by noting that the phonon peak is effectively shifted to
\beq
     \omega &=& \omega(q) -\frac{i}{2}\gamma_2(q,\omega(q))  . 
\eeq   
The peak, which is relatively narrow, again exhausts the f-sum rule, with contributions of the tail of $S$ to the sum rule being cancelled by renormalization corrections; see Appendix A.  The Beliaev process spreads what would be a sharp single phonon peak at $\omega=\omega(q)$ into the two phonon continuum.  

    The single phonon peak in $S(q,\omega)$ for anomalous dispersion does not cut off abruptly with decreasing $\omega$ at $\omega = \omega(q)$.  Rather
with $\omega$ increasing from zero, the first possible decay of a phonon of momentum $q$ is into two equal momentum phonons;  thus the minimum $\omega$ at which $\gamma_2$ is non-zero is $\omega_{min}= 2\omega(q/2)$, or
\beq
  \omega_{min} = \omega(q) - \frac34 s\zeta_A q^3.
\eeq
The structure of $S$ near $\omega=\omega(q)$ for anomalous as well as normal dispersion is shown in Fig.~\ref{Sscartoon}.

    The rate of excitation of the $^4$He via process (b) in Fig.~\ref{fig:dmtophonons} (denoted by the subscript $b$) is
\beq
   d\Gamma_b = n_\chi \left(\frac{2\pi a}{m_r} \right)^2   2\pi n_4  S_b(q,\omega) \frac{d^3q}{(2\pi)^3},  
\eeq
where 
\beq
     S_b(q,\omega) &=& \frac{q^2}{\pi m_4}\frac{\omega(q)\gamma_2(q,\omega)}{(\omega^2 - \omega(q)^2)^2 + \omega(q)^2\gamma_2(q,\omega)^2}
   \label{sgammab}
\eeq
is the structure function, as derived in Appendix A, with only the contribution from two-phonon states in $\gamma(q,\omega)$ included, and self-energy effects $\cal R$ in the denominator of Eq.~(\ref{sgamma}) neglected.     With $\gamma_2$ from Eq.~(\ref{xs})
we find that the relative half width of the single phonon peak at half height is $|\omega-\omega(q)|/\omega(q)\sim q^4/2\eta$, which is
$\sim 0.3 q^4$ with $q$ measured in inverse Angstroms; the peak is very narrow.    In the limit in which the width $\gamma_2$ goes to 0, $S_b(q,\omega)$ becomes simply the one-phonon structure function, Eq.~(\ref{sdelta}).  

   For normal dispersion, Eq.~(\ref{sgammab}), with Eq.~(\ref{gammau}), implies that for $\omega$ just above $\omega(q)$,
\beq
   S_b(q,\omega) \simeq \frac{q^2{\cal C}}{4\pi m_4\omega(q)}(\omega-\omega(q)),
\label{snorm} 
\eeq

\begin{figure}[h]
\includegraphics[width=8cm]{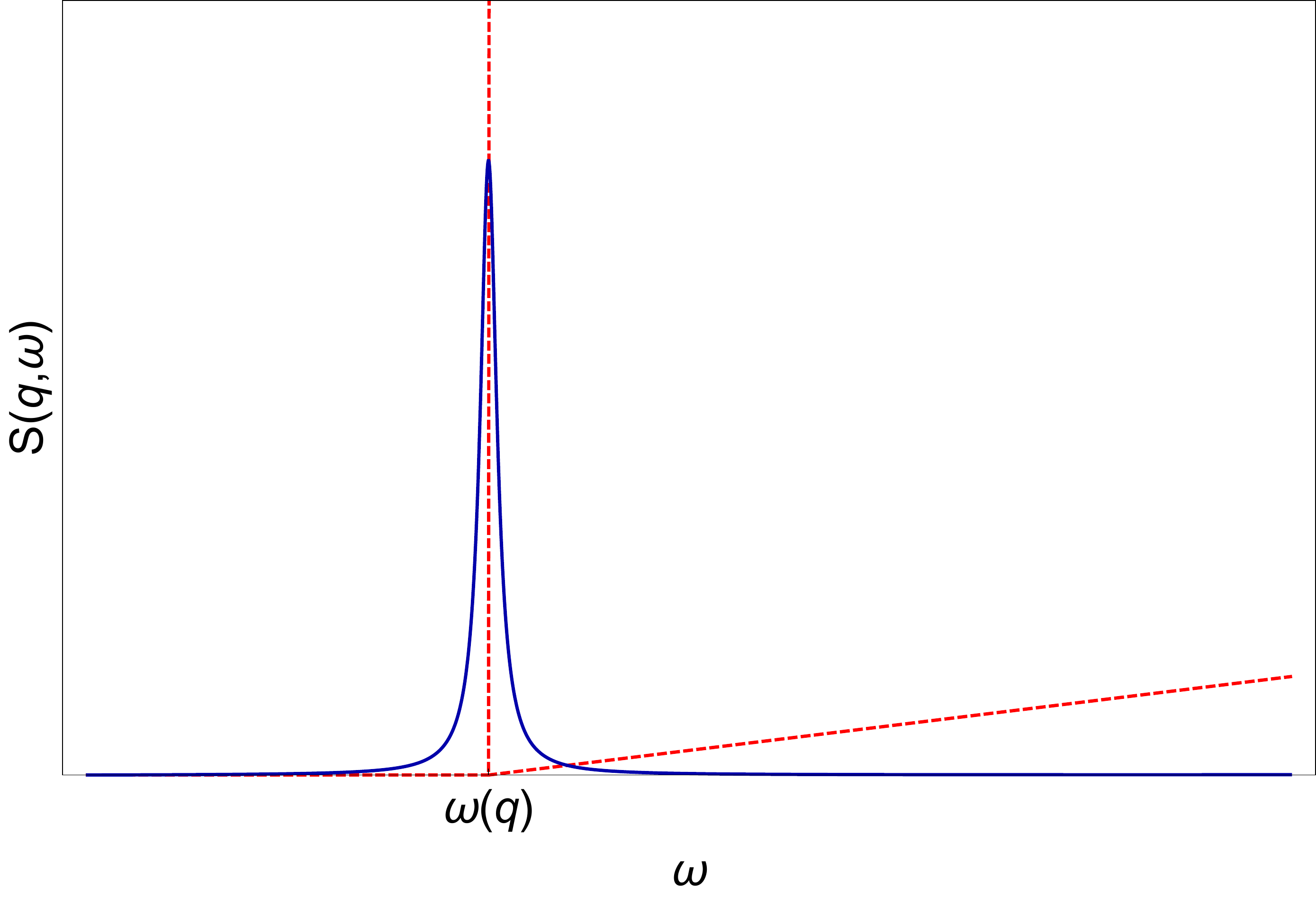}
\caption{Schematic behavior of $S(q,\omega)$ near $\omega(q)$.  At higher $q$, where the phonon dispersion is normal (\emph{dashed}), $S$ has a zero-width delta function at $\omega=\omega(q)$ and a multi-phonon contribution that is initially linearly increasing for $\omega > \omega(q)$, Eq.~(\ref{snorm}). At lower $q$, where the dispersion is anomalous (\emph{solid}), the delta function is spread out; see Eq.~(\ref{xs}).
} ad
\label{Sscartoon}
\end{figure}

   We note that for anomalous dispersion for small $q$, the integrand in Eq.~(\ref{gamma-d}) gives the relative probability of the initial phonon $q$ decaying to phonons with momenta $q_1$ and $q_2$.
The normalized probability for small $q$ is
\beq
   \frac{dP}{dx} = 30  x^2(1-x)^2,
         \label{Px}
\eeq
where $0\le x \le 1$.

\section{Two phonon emission}
\label{2phononsec}

  Production of multiple phonons proceeds by creation of an initial off-shell phonon which converts into multiphonon states, process (b) in Fig.~\ref{fig:dmtophonons}, or else by direct production of a multi-phonon state by the dark matter particle, process (c) in Fig.~\ref{fig:dmtophonons}.  These two processes are coherent.
Emission of two phonons via an intermediate single phonon has the amplitude  
\beq
   A_2^1&=& \frac{2\pi a}{m_\chi}  \langle \vec q_1\,, \vec q_2\,|\rho|0\rangle_{one\, phonon}
\label{a21}
\eeq 
times $\delta_{\vec q_1+\vec q_2\, ,\vec q\,} /\sqrt\Omega$,
where $|0\rangle$ is the $^4$He ground state, and
\beq
  \langle \vec q_1\,, \vec q_2\,|\rho|0\rangle_{one\, phonon} =  \hspace{96pt}\nonumber\\
\left(\frac{q^2n_4}{2m_4\omega(q)}\right)^{1/2} \frac{2\omega(q)\langle \vec q_1\,, \vec q_2\,|V|\vec q\,\rangle}{\omega^2-\omega(q)^2 +i\omega(q)\gamma(q,\omega)}.
\label{a210}
\eeq
The factor $\left(q^2n_4/2m_4\omega(q)\right)^{1/2}$ is the amplitude for the density operator to create a phonon of momentum $q$; the following  factor corresponds to the combination of the energy denominators $(\omega-\omega(q) +i\gamma(q,\omega)/2)^{-1} -(\omega+\omega(q) +i\gamma(q,\omega)/2)^{-1}$ describing the propagation of the phonon.  

    The direct production of two phonons by the dark matter particle, Fig.~\ref{fig:dmtophonons}c,  has amplitude
\beq
   A_2^d = \frac{2\pi a}{m_\chi} \langle \vec q_1\,, \vec q_2\,|\rho|0\rangle_{direct},
\eeq
times $\delta_{\vec q_1+\vec q_2\, ,\vec q\,} /\sqrt\Omega$.  
While the amplitude for the density operator to create two phonons directly is not well determined over the relevant range of momenta, its leading dependence at low $q$ is $\sim q^2$;  this result has previously been demonstrated for specific models \cite{zurek1,Caputo,Caputo2} but, as we show in Appendix C, it is a straightforward consequence of translational invariance.
Thus we write
\beq
\langle \vec q_1\,, \vec q_2\,|\rho|0\rangle_{direct} \equiv \frac{q^2}{m_4}\frac{\cal M}{sk_D}.
\label{rho2}
\eeq
We estimate that its strength is ${\cal M}\sim 1$ from sum-rule arguments in Appendix C.  

   The total matrix element for creating two phonons, the sum of (\ref{a210}) and (\ref{rho2}),  takes the form $ (q^2/m_4)\langle \vec q_1,\vec q_2|M|0\rangle$, where with linear phonon dispersion,
\beq
  \langle \vec q_1,\vec q_2|M|0\rangle &\equiv &\frac{s\sqrt{q_1q_2}(u+\nu)}
 {\omega^2-\omega(q)^2 +i\omega(q)\gamma_2(q,\omega)} +\frac{\cal M}{sk_D}.
 \nonumber\\ 
\eeq 
Near the resonance ($\omega\simeq \omega(q)$ in the denominator), the 
one-to-two phonon process is dominant.  However for $q_1, q_2 \sim k_D$ away from resonance the two amplitudes, $A_2^1$ and $A_2^d$, are comparable in magnitude. 

     The differential rate of two phonon emission per unit volume of $^4$He is 
\beq 
  d\Gamma_2& \simeq&  \frac{\pi \Gamma_0}{2 m_\chi k n_4} \int\frac{d^3 q_1}{(2\pi)^3}\frac{d^3 q_2}{(2\pi)^3}  \frac{q^4}{m_4^2}  |\langle \vec q_1,\vec q_2|M|0\rangle| ^2 \nonumber\\
   &&\hspace{58pt}\times 2\pi\delta(\omega-\omega(q_1)-\omega(q_2)), 
      \label{diffgamma-formal}
\eeq  
where  the factor 1/2  again compensates for double counting of phonon final states;
here  $\vec q = \vec q_1+\vec q_2$.       Comparing Eq.~(\ref{diffgamma-formal}) with Eq.~(\ref{gammaS}) we see that the dynamic structure factor for two-phonon states is
\beq
   S_2(q,\omega) &=&  \frac{3q^4}{8\pi k_D^3 m_4^2}  \int d^3q_1 d^3q_2 \delta\left(\omega-\omega(q_1)-\omega(q_2)\right) \nonumber \\
    &&\hspace{36  pt} \times\delta(\vec q - \vec q_1 - \vec q_2)   |\langle \vec q_1,\vec q_2|M|0\rangle| ^2 .
  \label{s2}
\eeq
 We first evaluate $S_2(q,\omega)$ for $\omega$ large compared with $\omega(q)$.   With the momentum delta function used to eliminate the $q_2$ integral, the $\cos\delta_1$ integral, as in Eq.~(\ref{intcosdelta1}), gives a factor $q_2/qq_1s$, so that
\beq
   S_2(q,\omega) &=&  \frac{3q^3}{4 k_D^3 m_4^2s}  \int_0^{\omega/s} dq_1 q_1q_2  |\langle \vec q_1,\vec q_2|M|0\rangle| ^2 . \nonumber\\
  \label{s2a}
\eeq

    The limits on the range of the integration over $q_1$, for $q \ll \omega/s$, are\footnote{In the published version of this paper \cite{he-dm} the $x$-integration was done in error from $x= 0$ to 1.  We thank Yoni Kahn for bringing the mistake in the published Eq.~(\ref{xint})  to our attention.} $  (\omega/s-q)/2$ and $(\omega/s+q)/2$.  Doing the $q_1$ integral with $q_1= \omega x$ and $q_2 = (1-x)\omega$ as above,  we find 
 \beq
S_2(q,\omega) =  \frac{3}{16}\frac{q^4 }{k_D^3 m_4^2s^3}  \left(\frac{(u-1)}{2} +\frac{\cal M \omega}{sk_D}\right)^2, 
  \label{xint}
\eeq
where we let $\nu \to -1$ for approximately back-to-back phonons.  This equation agrees with Eq.~(C3).   In the regime of $\omega$ that can produce two phonons, $S(q,\omega)$ is approximately a quadratic function of $\omega$.

  The rate of two phonon emission in this regime is given by Eq.~(9) with Eq.~(56) above, with the upper limit on the $\omega$ integral essentially $2sk_D$.  Thus
\beq
  \frac{d\Gamma_2}{dq}&&= \frac{\Gamma_0}{4} \frac{r q^5}{(k k_D m_4s)^2}   \nonumber\\
    &&\simeq  \frac{r q^3}{k_D^2m_4s}\frac{d\Gamma_1}{dq}. 
    \label{dgamma2dq}
\eeq 
Here $r={\cal M}^2 
    + (3/4){\cal M}(u-1)+(3/16)(u-1)^2$, which for the two allowed values of $\cal M$ is 0.77 for ${\cal M}=-1.47$ and 1.04 for ${\cal M}=0.25$. 
This estimate is consistent with the sum rule result, Eq.\ (23).

   Integrating over $q$ from 0 to $2k=2m_\chi v$ and averaging the velocity $v$ using the halo model as above, we find that the total zero-threshold two-phonon rate is given by
\beq
   \langle \Gamma_2(\omega>0)\rangle =  \frac{8r}{3} \frac{\sigma_{\chi_4}n_4\rho_\chi}{k_D^2(m_4s)^2} m_\chi^3 I,
\eeq 
where the velocity integral  $I=\int dv\, v^5 f(v)\simeq (427 {\rm km/s})^5$.
Numerically,
\beq
&& \langle \Gamma_2(\omega>0)\rangle \nonumber\\ &&\approx  1.0 \times 10^{- 8}{\rm cm}^{-3}{\rm s}^{-1}\, r\frac{\sigma_{\chi_4}}{10^{-40}{\rm cm}^2}
    \left(\frac{m_\chi}{1\,{\rm MeV} } \right)^3.
\eeq 
The two-phonon rate is less than one percent of the one-phonon rate for $m_\chi \simle 200 $~keV;
 the two averaged rates are approximately equal for $m_{\chi}\sim$ 0.9~MeV.

 \section{Phonon cascades in the anomalous dispersion regime} 
 \label{cones}

   We next describe the behavior of individual phonons produced in superfluid $^4$He by a dark matter particle.   Owing to anomalous dispersion, phonons up to the critical momentum, $q_c$, decay rapidly into two phonons, the Beliaev process.     The lifetime of a single on-shell phonon of momentum $q$ to decay into a pair of phonons is given by Eq.~(\ref{xs}), and the  corresponding phonon mean free path is given in Eq.~(\ref{mfp}).    As a consequence of this rapid decay, a phonon of momentum below $q_c$ will generate a cascade of lower momentum phonons, as illustrated in Fig.~\ref{fig:schematic}.    Appendix C gives an extended description of the decay of a single phonon into two.

\begin{figure}
\includegraphics[width = 0.45\textwidth]{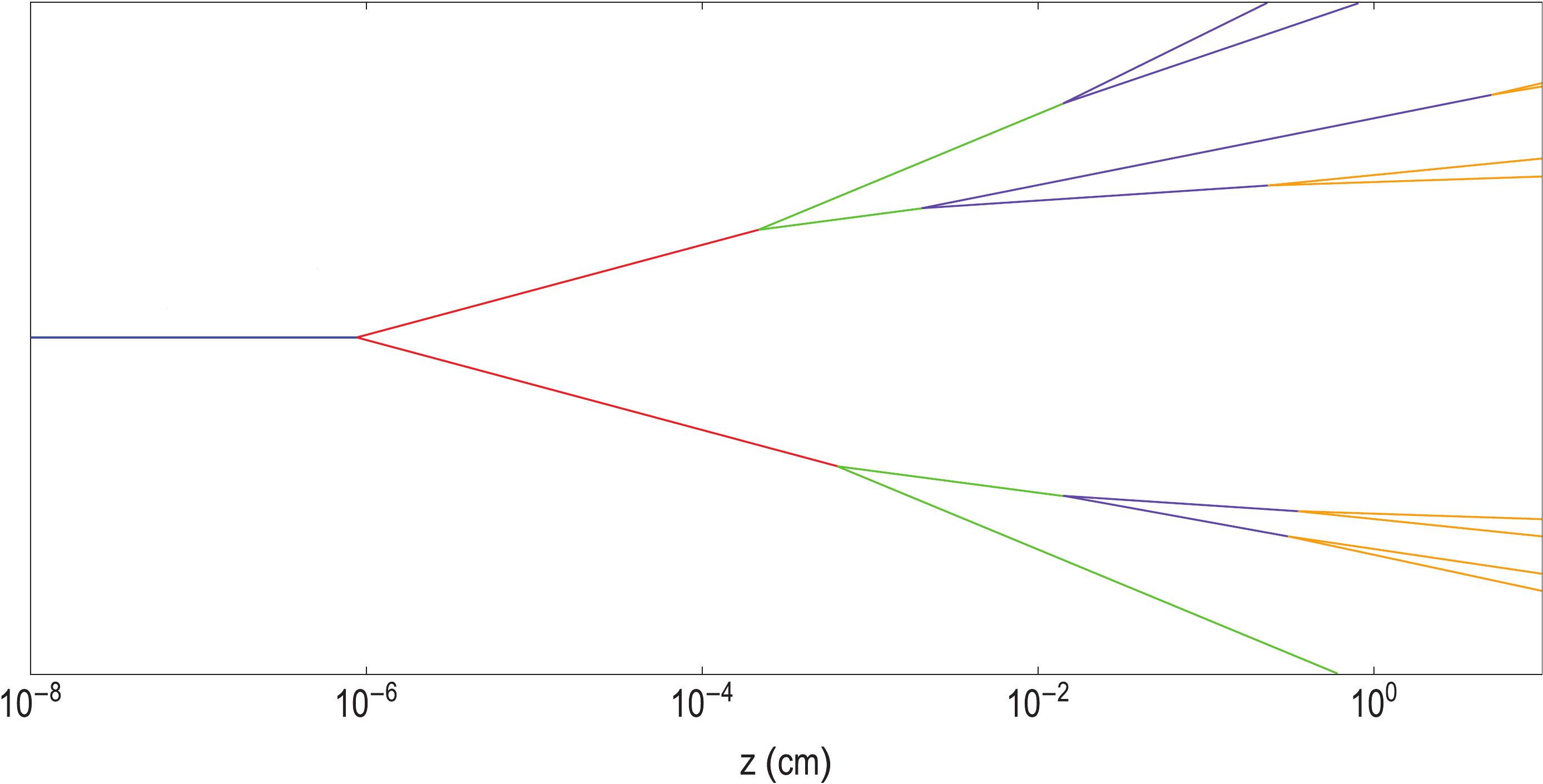}
\caption{Schematic drawing of a cascade of phonons in the anomalous dispersion region.  The initial phonon energy is 0.5 meV ($q$=0.32\AA$^{-1}$); the average phonon energy in the third (purple) generation is about 0.06 meV where the mean free path is of order cm.  Path lengths and energy splittings in the figure correspond to Eqs. (\ref{mfp}) and (\ref{Px}), respectively. The horizontal scale is logarithmic and the vertical scale is essentially the distance from the extension of the path of the initial phonon, divided by $z$.     For illustration, the paths are drawn as straight lines.}
\label{fig:schematic}
\end{figure}

    We first derive the opening angle in the three-phonon process,  $\vec q_1 \to \vec q_3 +\vec q_4$.  For simplicity, we present the calculation only in terms of the simplified anomalous dispersion relation $\omega(q) \simeq sq(1+\zeta_A q^2)$.    Since the deviations from linear dispersion in the anomalous dispersion regime are small, the angles $\delta_3$ and $\delta_4$ of the daughter phonons with respect to $\vec q_1$ are also small.  Using  $q_4^2 = (\vec q_1- \vec q_3\,)^2 =
q_1^2 + q_3^2 - 2q_1q_3\cos \delta_3$, in the energy conservation condition,  $\omega_3+\omega_4 = \omega_1$,
we find that to leading order (neglecting small terms $\sim \zeta_A q_1^2 \delta_3^2$, etc.),
\beq
  \delta_3^2 \simeq 6 \zeta_A  q_4^2, \quad \delta_4^2 \simeq 6 \zeta_A q_3^2.
  \label{psi}
\eeq

     The mean  $\langle\delta_3^2\rangle$ of the phonon $\vec q_3$ is given by the integral of (\ref{psi}) weighted by the probability (\ref{Px}) that the daughter phonon carries a fraction $x$ of the original phonon energy,
\beq
   \langle \delta_3^2\rangle &=& \int_0^1 dx\,6 \zeta_A q_1^2 (1-x)^2\,\frac{dP(x)}{dx} = \frac{12}{7} \zeta_A q_1^2. 
   \eeq
In a dark matter experiment, an initial phonon of momentum $\sim 0.1{\rm \angstrom}^{-1}$ in its first decay into two phonons would lead to an rms opening angle $\simeq 7.9 ^\circ$ between the initial phonon and each daughter.\footnote{The present calculation is roughly consistent with the measurement of Wyatt et al.~\cite{wyatt} in which they applied a collimated heat pulse at temperature $T_p$ (not the ambient helium temperature) and measured the opening angle of the phonon cone produced,
finding angles of order 8$^\circ$ for a thermal distribution of initial phonons at $T_p$ = 2.2K.}

    To see the general
structure of the cascade of an initial phonon of momentum $q_0$ large compared with momentum of thermal phonons, $\sim T/s$, at the ambient temperature
$T$, we make the simplifying assumption that in each decay process a single phonon of momentum $q_i$ divides into two
phonons each of momentum $q_i/2$. [This is a reasonable approximation given the peaking of the probability (\ref{Px}) about $x$=1/2; the mean square deviation from 1/2 is $\langle(x-1/2)^2\rangle$ =1/28. The phonons will be at a slight angle with respect to each other, as calculated above.  
Were the mean free path a constant, $\ell_0$,  the number of phonons, $N(z)$, present at a distance $z$ from the initial phonon production point would be $N(z) \simeq 2^{z/\ell_0}$.
The mean free path depends however on $q$ as $\ell(z) = s\tau_{3} = \eta/q^5$, and thus we can more generally write
\beq
   \frac{dN(z)}{dz} =\frac{N(z)}{\ell(z)} \ln 2  = \frac{\ln 2}{\eta}\left(\frac{q_0}{N(z)}\right)^5 N(z),
   \label{dndz}
\eeq
since the initial phonon energy becomes spread among $N$ phonons, with
average phonon momentum $q_0/N$.  The solution of Eq.~(\ref{dndz}) is
\beq
   N(z) = \left( 1+(5\ln2) \frac{z}{\ell_0}\right)^{1/5},
   \label{N}
\eeq
where $\ell_0 = \eta/q_0^5$ is the mean free path of the initial phonon.   For $q_0 = 0.5$ \angstrom$^{-1}$ (or phonon energy 9 K), we have $\ell_0 \sim 53 $ \angstrom, and in a cascade of length 30 cm, the final number of phonons in the cascade is $N \sim 42$.   The mean phonon energy is $\sim 0.21$ K, an order of magnitude larger than the expected ambient temperature of $\sim 10$mK.     The phonon mean free path exceeds the radius, $R$, of the helium container, only for $q\lesssim 0.014{\rm\,\AA}^{-1}(30\,{\rm cm}/R)^{1/5}$. 

   We next estimate the widening of the cascade with subsequent decays, again with the simplifying assumption that phonons split only into pairs of equal energy.   We keep only the $\zeta_A$ term in the dispersion.  Then
after $n$ splittings, the angle $\delta_n$ of an $n$th generation phonon with respect to its progenitor phonon of momentum $\vec q_{n-1}$ is
\beq
    \delta_n^2 \simeq \frac32\zeta_A \left(\frac{q}{2^{n-1}}\right)^2,
   \label{anglen}
\eeq
and its momentum is given by
\beq
    \vec q_n =  \frac12\left(\vec q_{n-1} +\delta_n q_{n-1} \hat{m}_n\right),
\eeq
where $\hat m_n$ is a unit vector orthogonal to $\vec q_{n-1}$.
The opening angle measured with respect to the initial phonon $\vec q_0$ is then given by  $ \cos \theta_n\equiv \hat q_n\cdot \hat q_0$,
so that to leading order 
\beq
  \theta_n^2 =\delta_n^2 + \theta_{n-1}^2 -2 \delta_n \hat q_0 \cdot \hat m_n.
\eeq 
When averaged over $\hat m_n$ the last term goes away, and using Eq.~(\ref{anglen}) at $q_1=q_0/2$ we find
\beq
   \theta_n^2 =\frac32\zeta_A \left(\frac{q_0}{2^{n-1}}\right)^2 + \theta_{n-1}^2,
\eeq
a recursive relation with the solution
\beq
   \theta_{n}^2 = 2 \zeta_A q_0^2\left(1-\frac{1}{4^n} \right). 
   \label{thetacone}
\eeq
This simple estimate implies that the opening angle of the cone increases with subsequent phonon decays only by a factor $\simle 2/\sqrt 3$;
the basic physics is that the smaller the momentum of the phonons the smaller the opening angle in the decay.

    As is shown in Fig.~\ref{fig:mfpTheta}, which assumes that the daughter phonons have equal energy, the phonon mean free path in the anomalous dispersion region, Eq.~(\ref{mfp}), increases rapidly as $2^{5n}$, where $n$ is the generation index as the phonons split into pairs; at the same time, the angle between the daughter phonons decreases as $2^{n}$.  As a result the number of phonons in the shower depends strongly on the initial phonon momentum and increases only slowly after travelling distances of order 1 cm in the detector, as shown in Fig.~\ref{fig:NPhonons}.

\begin{figure}[t]
\includegraphics[width=9.5cm]{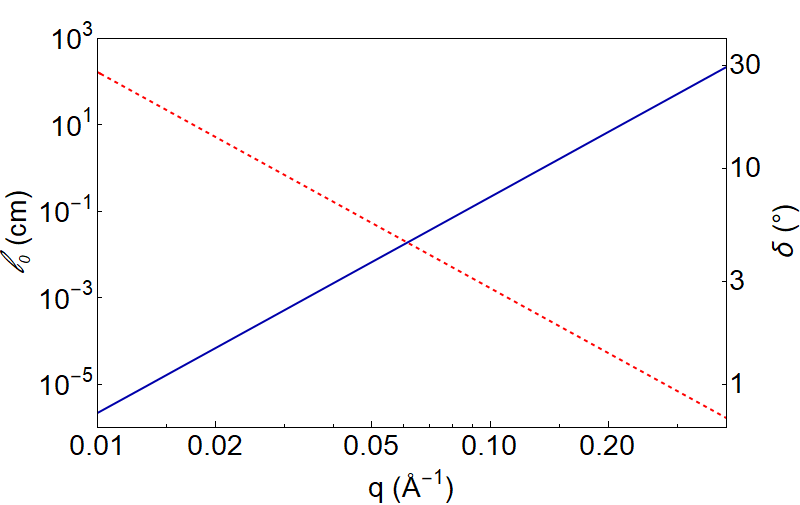}
\caption{The mean free path, $\ell$ (\emph{dashed, left axis}), and the rms angle in degrees of the daughter phonons, $\delta$ (\emph{solid, right axis}), as a function of phonon
momentum, $q$ for the case of phonons split evenly in energy. Note that the mean free path varies rapidly, from a few hundred $\rm{\angstrom}$ at $q = 0.4
\rm{\angstrom}^{-1}$ (somewhat below the maximum momentum that can decay into two phonons) to $\sim$ 2~cm at $q = 0.4/2^4 = 0.025\rm{ \angstrom}^{-1}$.}

\label{fig:mfpTheta}
\end{figure}

\begin{figure}[t]
\includegraphics[width=8.5cm]{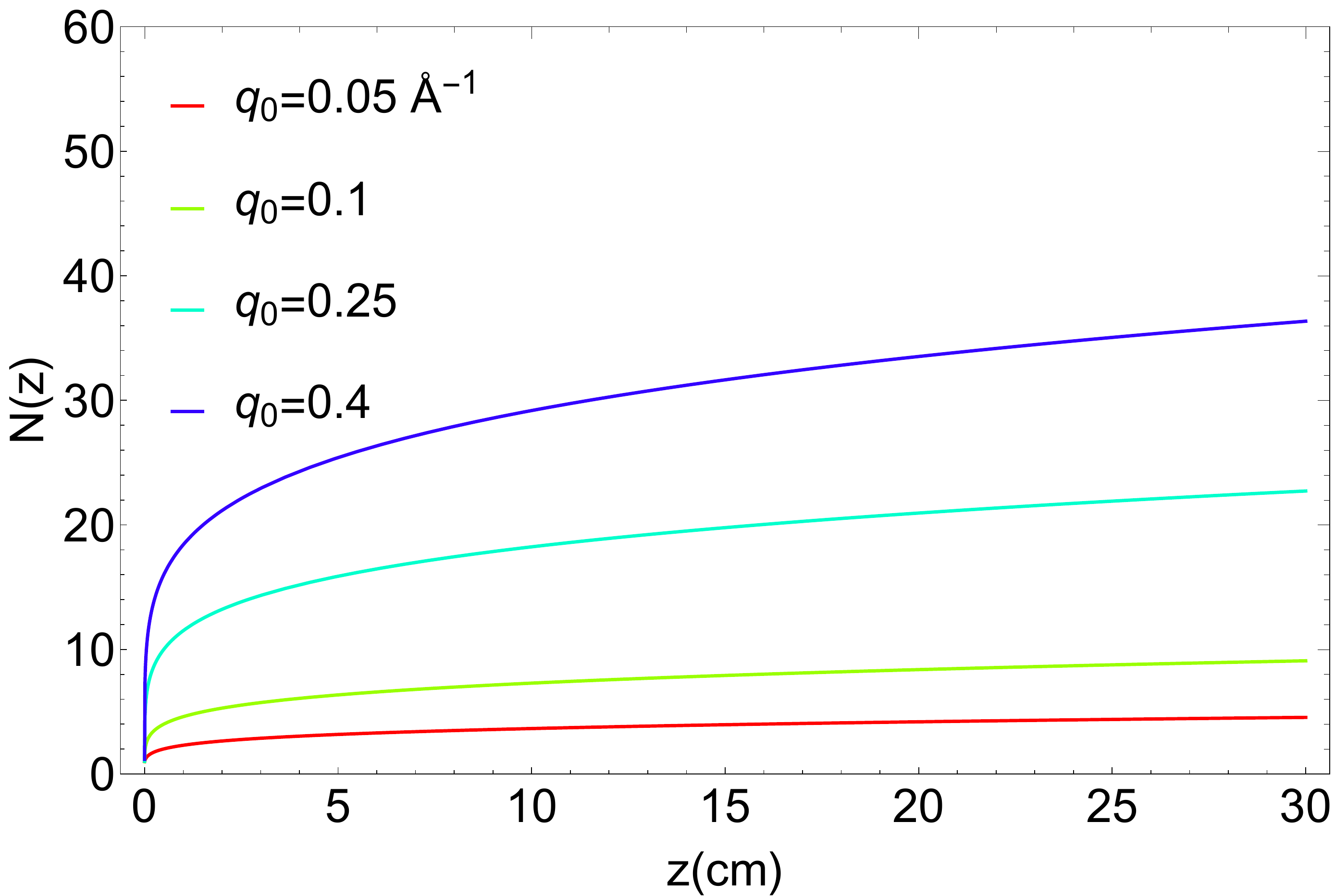}
\caption{The number of phonons, $N(z)$, produced as a function of path length, $z$, for a number of initial momenta, $q_0$, and for the case of phonons split evenly in energy.  The number of phonons in the shower increases rapidly for the first few cm of the path and thereafter increases relatively slowly.}
\label{fig:NPhonons}
\end{figure}

   A detector that measures simply the total energy of an excitation at the container wall does not provide information on where the excitation was produced.   The geometry of a phonon cascade, however, carries with it such information.   The characteristic signature of a phonon cascade is that its energy deposition in detectors on the surface of the helium will be elliptically shaped.  This signature provides discrimination against background events in the detectors.  The orientation angle of the major axis of the ellipse, together with the ratio of the minor to major axes, tells one the direction cosines of the original velocity vector with respect to the surface.   The size of the minor axis tells one the opening angle of the phonon cascade, from which one can deduce the distance from the center of the ellipse to the original vertex, see Eq.~(\ref{thetacone}). A typical pattern of energy deposition on the surface of a sphere is shown in Fig.~\ref{ellipse}.  The total energy deposition is a measure of the energy transferred by the dark matter particle.   With sufficient detector sensitivity on the surface one can take advantage of the anomalous dispersion to pin down the event.

 The energy of the initial phonon equals the total energy of
phonons arriving at the detector, and therefore a measurement of the
latter would determine the energy as well as the magnitude of the
momentum of the initial phonon.   A natural question to ask is whether, for
a initial phonon produced with momentum in the anomalous region of the dispersion curve, the
spatial distribution of secondary phonons arriving at the detector can provide
information about the origin of the initial phonon.  Each such event
produces a pair of daughter phonons with a relatively wide opening angle and a random
azimuthal orientation; subsequent splittings of the phonons in the cascade proceed with successively
smaller opening angles.   The overall elliptic pattern that an event would produce at the surface (schematically shown in Fig.~\ref{ellipse} for an event with many phonons) contains information
about the location of the original event, but it is a quantitative question as to whether
the uncertainties obscure the significance of such location information.  In principle, one
would like to do a simulation, starting with a candidate dark matter mass, to determine the extent to which 
the event patterns at the detector surface can constrain the
extracted mass, a task we leave for future work.

\begin{figure}[t]
\includegraphics[width=10cm]{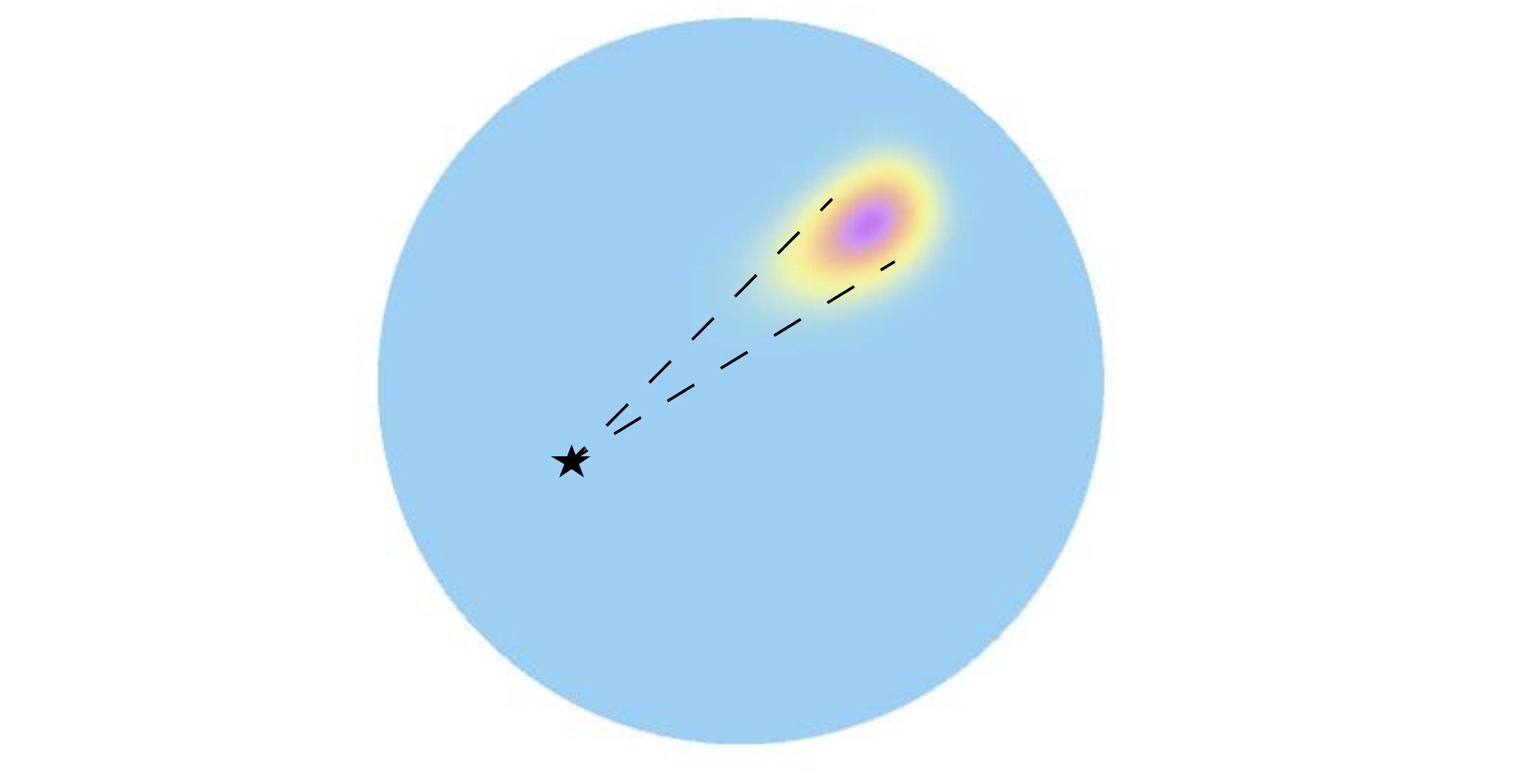}
\caption{A schematic illustration of the pattern of elliptical energy deposition on the surface of a spherical container of helium resulting from a phonon cascade.  The dashed lines symbolically indicate the phonon cone, which expands according to Eq.~(\ref{thetacone}), and the vertex of the interaction of the dark matter with the helium.}
\label{ellipse}
\end{figure}

\section{conclusions}

     We have discussed here the processes that will occur when a
low-mass dark matter particle scatters from liquid $^4$He.  
As we have shown, the total strength of the
excitation function for states with two or more phonons is tightly
constrained by sum-rule arguments; these become increasingly severe
as the momentum transfer to the helium declines, and thus at lower 
dark matter particle masses.  For
dark matter masses $\lesssim$ 1.2~MeV, the most probable 
outcome is the creation of a single phonon, and this process becomes overwhelmingly dominant at keV masses.
Moreover, at liquid pressures less than $\sim 18$~
bar, this situation is made more interesting by the fact that a single phonon can
decay into a cascade of lower-energy phonons.
Consequently, the distinction between single-phonon and 
multi-phonon scattering is blurred;
a two-phonon state, for example, can be reached from the ground state of $^4$He
by either direct creation of two phonons or by creation of a single phonon
which subsequently decays, and the net rate of production of
phonon pairs is a coherent superposition of these two processes.
As these arguments indicate, and as discussed in ~\cite{zurek,zurek1}, two-phonon processes
that can produce phonons in the several-meV range grow in relevance 
for dark matter masses $\gtrsim$~MeV. 
Our discussion complements that of Zurek and collaborators by exploring 
the processes relevant at lower momentum transfers and nearer to the dispersion curve.

Our findings present both experimental challenges and
opportunities.    Although a thorough consideration of detection techniques is beyond the scope of the paper, we address briefly two key aspects relevant to sub-MeV dark matter.    In this mass regime, the bulk of the events will appear in the form of creation of single
phonons, pointing to the need to detect phonons of energy 1 meV or lower. 
Above $\omega(q_c)\sim$ 0.7 meV such phonons travel ballistically, and can be detected 
via surface evaporation or direct absorption by bolometric detectors~\cite{hertel}.
This technique could be extended to lower energies by using a pressurized helium vessel 
to suppress anomalous dispersion.

  At low pressures 
phonons of energy below $\sim$ 0.7 meV will decay into a cascade of even lower energy phonons, so it becomes
necessary to develop methods to detect such showers. 
Such decays will occur even if a pair of such phonons is created initially.
If detectable, the resulting shower would contain information about the direction and 
magnitude of the total momentum imparted to the helium, in much the same way as the
direction of the incoming primary that creates a cosmic ray shower is
determined.  
That said, even with sensitive detectors such information would be very difficult to obtain. The probability of a phonon escaping from liquid He and depositing its
energy in a detector is substantially reduced by Kapitza resistance: Tanatarov et al.~\cite{tanatarov10} note $\sim$1.5\% transmission at best, and then only at normal incidence.  

To  assess detection possibilities, it is instructive to consider detection of the energy deposited by low energy {\it photons}.  In an example from this rapidly developing field, the authors of Ref.~\cite{echternach17}
used a quantum capacitance detector to measure, with high efficiency, single
photons from a 5~K blackbody source which has a most probable photon energy of about 2~meV.  
Similar techniques may point the way to direct detection of low energy phonons.
Although current progress is encouraging, broadly speaking, the use of superfluid $^4$He detectors for dark matter detection awaits further development of detection techniques.   With improved detection capabilities, superfluid $^4$He may have the potential not only to record events produced by low mass dark matter particles, but also to pin down the particle mass and interaction cross section with baryonic matter.

\section*{Acknowledgements}

This research was supported in part by NSF Grants PHY1714042 and PHY18-22502.  The work of JS is supported by DOE CAREER Grant DE-SC0017840.   Author GB is grateful to the Aspen Center for Physics, supported by NSF Grant PHY1607611,  and to the Niels Bohr International Academy,  where part of this research was done.

\appendix

\section{Relation of the dynamical structure function to the $\mathbf{^4}$He density correlation function}
\label{AppendixA}

    We recall here the relation of $S(q,\omega)$ for real frequency $\omega$ to the density-density correlation function in the complex frequency plane.  In general, at temperature $T$ the density correlations $\langle\rho \rho\rangle (q,z)$, where in standard condensed matter notation $z$ is the complex frequency,  are defined by \cite{KB}
\beq
    &&  \langle \rho(\vec r, t) \rho(\vec r\,', t')\rangle =\nonumber\\ &&iT\sum_{n=-\infty}^\infty e^{-iz_n(t-t'))}\int\frac{d^3q}{(2\pi)^3} e^{i\vec q\cdot(\vec r-\vec r\,')} \langle\rho \rho\rangle (q,z_n), \nonumber\\
\eeq
where $z_n = 2\pi i Tn$ are the Matsubara frequencies.  Then  $\langle\rho \rho\rangle (q,z)$, the analytic continuation of $\langle\rho \rho\rangle (q,z_n)$ to the complex frequency $z$ plane, has the form,
\beq
    \langle\rho \rho\rangle (q,z) = \frac{n_4q^2/m_4}{z^2 -\Pi(q,z)},
    \label{rhorho0}
\eeq  
 and is given in terms of the structure function by
\beq
   \langle\rho\rho\rangle(q,z) &=& n_4\int_{-\infty}^\infty d\omega \frac{ S(q,\omega)-S(q,-\omega)}{z-\omega}
   \label{rhorho1} \\
    &=& 2n_4\int_{-\infty}^\infty d\omega \frac{ \omega S(q,\omega)}{z^2-\omega^2}.
   \label{rhorho}
\eeq
 Comparison of the large $z$ limit of Eqs.~(\ref{rhorho}) and (\ref{rhorho0}) yields the f-sum rule, Eq.~(\ref{sumrule}). At zero temperature $S$ vanishes for negative frequency.   The following Appendix discusses $S(q,\omega)$ at finite temperature.
 
We write $\Pi$ in terms of its real and imaginary parts, 
$
  \Pi(q,\omega+i\epsilon) =  \Re\Pi  + i\Im \Pi,
$
where $+i\epsilon$, with $\epsilon\to0$, indicates a limit to the real axis from the upper half plane.  In this limit, $\Im \Pi \le 0$, for $\omega \ge 0$.  The imaginary parts of Eqs.~(\ref{rhorho1}) and (\ref{rhorho0}) then imply
\beq
   S(q,\omega) &=& \frac{q^2}{\pi m_4}\frac{-\Im\Pi}{(\omega^2 -\Re\Pi)^2 + \Im\Pi^2}.
   \label{sgamma}
\eeq

   At low $q$ where single phonon excitations dominate, $\Pi$ has the structure,
\beq
   \Pi(q,\omega) = \omega(q)^2 +{\cal R}(q,\omega)-i\omega(q)\gamma(q,\omega),
\eeq
where $\cal R$ is real; then $S(q,\omega)$ for $\omega \ge 0$ becomes,
\beq
   S(q,\omega) &=& \frac{q^2}{\pi m_4}\frac{\omega(q)\gamma(q,\omega)}{(\omega^2 - \omega(q)^2-{\cal R})^2 + \omega(q)^2\gamma(q,\omega)^2}.
   \label{sgamma}
\nonumber\\
\eeq
The function $\gamma(q,\omega)$, which is non-negative, determines the damping rate of the density excitations.  With $\cal R$ in the denominator of Eq.~(\ref{sgamma}) neglected, and $\gamma \to \gamma_2$ in the numerator,  Eq.~(\ref{sgamma}), derived from the structure of the density-density correlation function, reduces to the ``one-to-two'' phonon contribution to Eq.~(\ref{sgammab}) with $\gamma_2$ given by Eq.~(\ref{gamma}).

   In the limit $\gamma\to 0$, the structure function for $\omega \ge 0$ reduces to 
\beq
   S(q,\omega) &=& \frac{q^2}{m_4}\delta(\omega^2 - \omega(q)^2-{\cal R}(q,\omega))^2 )\nonumber\\ 
   &=& \frac{q^2}{2m_4\omega(q)_R}\frac{\delta(\omega - \omega(q)_R )}{\left(1-\partial {\cal R}/\partial\omega^2|_{\omega=\omega(q)_R}\right)}.
   \label{sgamma1}
\eeq
The term ${\cal R}(q,\omega)$ serves to renormalize the excitation energy from $\omega(q)$ to the solution, $\omega(q)_R$, of $\omega^2 = \omega(q)^2+{\cal R}(q,\omega)$.  
We neglect such renormalization effects throughout.   The factor $Z(q)$ in the decomposition (\ref{eq:sqomega}) is the coefficient of the $\delta$ function in Eq.~(\ref{sgamma1}).   Renormalization of the excitation energy and the decrease of the contribution of the single phonon peak to the f-sum rule are intimately related.
In the limit $\gamma\to 0$  with $\cal R$ neglected, the structure function reduces to the single phonon result, Eq.~(\ref{sdelta}).

\section{$S(q,\omega)$ in HeII at finite temperature}
\label{appendixB}

   Since in general at finite $T$,  $S_T(q,-\omega) = e^{-\omega/T}S_T(q,\omega)$ we can write
\beq
   S_T(q,\omega) =  \left(1+n(\omega)\right)B(q,\omega),
\eeq
where $n(\omega) = 1/\left(e^{\omega/T} -1\right)$
is the Bose occupation factor at energy $\omega$,
and $B(q,\omega)$ is odd in $\omega$ and non-negative for $\omega > 0$.  

   The f-sum rule becomes
\beq
    \int_{-\infty}^\infty d\omega\, \omega \left(1+ n(\omega)\right) B(q,\omega) \nonumber\\ = \int_0^\infty d\omega \, \omega  B(q,\omega)
  = \frac{q^2}{2m_4},
\eeq  
while the static structure function becomes
\beq
   S_T(q) = \int_{-\infty}^\infty d\omega\, \left(1+ n(\omega)\right) B(q,\omega) \nonumber\\ = \int_0^\infty d\omega\, \left(1+ 2n(\omega)\right) B(q,\omega). 
\eeq

   Separating out the single phonon contribution to $B$, we write for $\omega>0$,
\beq
    B(q,\omega) = Z(q)\delta(\omega-\omega(q)) + B_M(q,\omega),
\eeq   
where $B_M \ge 0$ is the multi-excitation contribution.   Then 
\beq
   S_T(q,\omega) = (1+n(\omega(q)) Z(q) \delta(\omega-\omega(q))) \nonumber\\ + (1+n(\omega) )  B_M(q,\omega);
\eeq
the weight of the phonon pole at finite temperature is
\beq
   Z_T(q) = (1+n(\omega(q)) Z(q).
   \label{zt}
\eeq

   The f-sum rule then implies that the multi-excitation contribution obeys
\beq
   \int_0^\infty d\omega \, \omega  B_M(q,\omega)  = \frac{q^2}{2m_4} - \omega(q)Z(q).
   \label{bmT}
\eeq
In addition,
\beq
  S_T(q) &=& \left(1+2n(\omega(q))\right)Z(q) \nonumber\\ &&+ \int_0^\infty d\omega\, \left(1+ 2n(\omega)\right) B_M(q,\omega). 
  \label{sT}
\eeq

  This equation enables us to place bounds on $S_T(q)$.   Since $\left(1+ 2n(\omega)\right)$ is a decreasing function of $\omega$,
and the support of $B_M(q,\omega)$ is essentially for $\omega \ge \omega(q)$, we have, with Eq.~(\ref{bmT}),
\beq
 && \int_0^\infty d\omega\, \left(1+ 2n(\omega)\right) B_M(q,\omega)\nonumber\\ &&< \left(1+ 2n(\omega(q))\right) \int_0^\infty d\omega B_M(q,\omega) \frac{\omega}{\omega{(q)}}\nonumber\\
 &&=\left(1+ 2n(\omega(q))\right)\left(\frac{q^2}{2m_4\omega(q)} - Z(q)\right).
\eeq
Thus from Eq.~(\ref{sT}) we have
\beq
   S_T(q) \le \frac{q^2}{2m_4\omega(q)} \left(1+ 2n(\omega(q))\right),
\eeq
while if we neglect the multi-excitation contribution to $S(q)$ we see that
\beq
   S_T(q) \ge  Z(q) \left[1+2n(\omega(q)\right].
\eeq
Altogether,
\beq
   Z(q) \le \frac{S_T(q)}{1+2n(\omega(q)} \le \frac{q^2}{2m_4\omega(q)}.
   \label{bound}
\eeq

 Liu and Woo \cite{liu-woo} give the expansions at $T=0$:
\beq
   S(q) &=& \frac{x}{2}(1 - 1.63x^2 + 1.42x^3 + 0.51x^4)\\
Z(q) &=& \frac{x}{2}(1 - 1.63x^2 - 0.78x^3 + 0.51x^4 - 2.46x^5), \nonumber
\label{expansions}
\eeq
where $x= q/m_4s$.  

   For small $q$  Eq.~(\ref{bound}) becomes
\beq
   1-z_2\left(\frac{q}{m_4s}\right)^2 \le \frac{S(q)}{\left(1+2n(\omega(q))\right)(q/2m_4s)} \le 2, 
\eeq
with $z_2 \simeq 1.63$.  Then to order $x^4$,
\beq
  S(q) -Z(q) = 1.1 \left(\frac{q}{m_4s}\right)^3,
\eeq
which, to within the error bars, is consistent with  the $Z(q)$ data from Cowley and Woods lying below that for $S(q)$ given by Robkoff and Hallock, as seen in Fig.~\ref{S(q)new}.

  We note furthermore that the single phonon contribution to  $S_T(q)$  at finite temperature is
$(1+2n(\omega(q)))Z(q)$,
while the weight of the pole is
$
   Z_T(q) =(1+n(\omega(q)))Z(q),
$
where $Z(q)$ is essentially the zero temperature weight of the pole.   The finite temperature effects on $S_T(q)$ are twice as large
as those on $Z_T(q)$.

\section{Matrix element for production of two phonons}
\label{AppendixC}

    We review how the matrix element for the two phonon process in Fig.~\ref{fig:dmtophonons}c  is constrained, for small momentum transfers, by the particle number conservation law \cite{NozieresPines}.  With the basic interaction between a dark matter particle and the helium atoms represented by a contact interaction, Eq.~(\ref{vchi4}), the $^4$He part of the matrix element is proportional to the Fourier transform $\rho_{-{\vec q}}$ of the density operator.  Particle number conservation, $
\partial \rho_{-\vec q}/\partial t- i\vec q \cdot {\vec j}_{-\vec q}=0$,
where $\rho_{-{\vec q}}$ and ${\vec j}_{-\vec q}$ are the Fourier transforms of the number and number current density operators,
implies that the matrix element of $\rho$ between the initial and final states obeys
\beq
\langle f|\rho_{-\vec q}\,|0\rangle=\frac{\vec q \cdot\langle f|\vec j_{-\vec q}\,|0\rangle}{E_f-E_0}.
\eeq
The $E$'s are the energies of the initial and final states.
Since the system is translationally invariant, the energy eigenstates may be also be taken to be eigenstates of the total momentum, which implies that, for $\vec q\to 0$,  ${\vec j}_{-\vec q}$ has no off-diagonal  matrix elements. 
Thus one expects that for small $q$, and for different initial and final state energies,
$
\langle f |{\vec j}_{-\vec q}|0\rangle \propto q^\alpha\hat q,
$
with $\alpha>0$ and 
\beq
\langle f|\rho_{-\vec q}\,|0\rangle\sim \frac{q^{1+\alpha}}{E_f-E_0},
\label{2phononnonresonant}
\eeq
when the magnitudes of
the phonon momenta in the final state are held fixed.  As $q\to0$ the energy denominator remains finite for two (or more) phonon excitations.
Analyticity in $q$  implies that the smallest value of $\alpha$ is unity, and so $\langle f|\rho_{-\vec q}\,|0\rangle \propto q^2$, as in Eq.~(\ref{rho2}).
By contrast, the matrix element of the current operator for creation of a single phonon is anomalous, varying as $q^{1/2}$, as explained in Ref. \cite{NozieresPines}.

   We now estimate the dimensionless matrix element ${\cal M}$ using the sum rule result, Eq.~(\ref{sumrulerates}).    We let $q\to 0$ in the integral in Eq.~(\ref{s2}),
so that $\vec q_2 = - \vec q_1$, i.e., the phonons are back-to-back; then, in agreement with Eq.~(\ref{xint}),
\beq
   &&S_2(q\to 0,\omega) =  \nonumber \\
   &&\frac{3q^4}{8\pi k_D^3 m_4^2}  \int d^3q_1\delta(\omega-2\omega(q_1))
   |\langle \vec q_1,-\vec q_1|M|0\rangle| ^2 \nonumber\\
   &=& \frac{3q^4}{16 k_D^3 m_4^2 s^3} \left( \frac12(u-1)+\frac{{\cal M}\omega}{sk_D} \right)^2.
     \label{s21}
\eeq
with the factor $(u-1)$ for back-to-back phonons.

     Including phonons up to momenta $sk_D$, we find the contribution to the f-sum rule from $S_2(q,\omega)$ for small $q$.
\beq
  &&  \int_0^{2sk_D} d\omega\,\omega S_2(q,\omega) \nonumber\\   
   &&  =\frac{3q^4}{4k_D m_4^2 s}  \left( \frac18(u-1)^2+ \frac23(u-1){\cal M}+{\cal M}^2\right). \nonumber\\
\eeq
Comparing with Eq.~(\ref{smsum}) we have
\beq
 \frac18(u-1)^2+ \frac23(u-1){\cal M}+{\cal M}^2 = \frac23 z_2 \frac{k_D}{m_4 s};
\label{m}
\eeq
numerically, ${\cal M}$ = -1.47 or +0.25.    Using Eq.~(\ref{m}), we find that the combination of terms, ${\cal M}^2 
    + (3\pi/16){\cal M}(u-1)+(u-1)^2/10$ in Eq.~(\ref{dgamma2dq}) is 0.7 or 0.9.

\end{document}